\documentclass[fleqn,usenatbib]{mnras}

\usepackage{newtxtext,newtxmath}

\usepackage[T1]{fontenc}

\DeclareRobustCommand{\VAN}[3]{#2}
\let\VANthebibliography\thebibliography
\def\thebibliography{\DeclareRobustCommand{\VAN}[3]{##3}\VANthebibliography}

\usepackage[mathscr]{euscript}
\usepackage{graphicx}	
\usepackage{amsmath}
\usepackage[dvipsnames]{xcolor}
\usepackage[nice]{nicefrac}
\newcommand{\orcid}[1]{\href{https://orcid.org/#1}{\includegraphics[width=8pt]{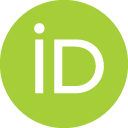}}}
\usepackage{subcaption}
\usepackage{caption}
\usepackage{tabularx}
\usepackage{booktabs}
\usepackage{amsfonts}
\usepackage{mathrsfs}

\title[GW190814 AGN]{Not all roads lead to merger: AGN disc properties influence the interactions of highly unequal mass black holes}

\author[Moncrieff et al.]{
Jordan W. N. Moncrieff$^{1,2}$\thanks{E-mail: jordan.moncrieff@research.uwa.edu.au}
\orcid{0009-0009-1028-5381}, 
Evgeni Grishin$^{2,3}$\thanks{E-mail: evgeni.grishin@monash.edu.au} 
\orcid{0000-0001-7113-723X}, %
Alessandro A. Trani$^{4,5}$
\orcid{0000-0001-5371-3432}, 
Fiona H. Panther$^{1,2}$
\orcid{0000-0002-2618-5627},
\newauthor
and Olga Pietrosanti$^{6}$
\\
$^{1}$ Department of Physics, University of Western Australia, Crawley WA 6009, Australia \\
$^{2}$OzGrav: Australian Research Council Centre of Excellence for Gravitational Wave Discovery, Australia\\
$^{3}$School of Physics and Astronomy, Monash University, Clayton, VIC 3800, Australia\\
$^4$Departamento de Astronom\'ia, Facultad Ciencias F\'isicas y Matem\'aticas, Universidad de Concepci\'on, 4030000, Concepci\'on, Chile\\
$^5$National Institute for Nuclear Physics – INFN, Sezione di Trieste, I-34127, Trieste, Italy\\
$^6$International School for Advanced Studies - SISSA, Via Bonomea 265, I-34136 Trieste, Italy
}

\date{Accepted XXX. Received YYY; in original form ZZZ}

\pubyear{2025}

\begin{document}
\label{firstpage}
\pagerange{\pageref{firstpage}--\pageref{lastpage}}
\maketitle

\begin{abstract}
As the number of gravitational-wave detections of black hole binaries grows, so does the diversity of proposed formation channels. The growing sample of systems with highly unequal masses, such as GW190814 with $m_1=23.2\,M_{\odot}$ and $m_2=2.59\,M_{\odot}$ -- corresponding to a mass ratio $q=0.112$ -- cannot be readily explained by isolated binary evolution and may originate through dynamical assembly in an active galactic nucleus (AGN). We investigate AGN discs capable of producing GW190814-like mergers using \texttt{pAGN} to model self-consistent AGN torques, coupled with \texttt{TSUNAMI}, a regularised N-body code including post-Newtonian terms up to 3.5 order. Suites of N-body simulations reveal possible outcomes of binary capture and merger, mean-motion resonance interactions, and other novel dynamical pathways. We develop analytical models linking the branching ratios of captures and mergers to local disc properties, applicable to black hole populations across all mass ratios. Capture probability is primarily governed by $\mathscr{B}$, the ratio of libration time to resonance-width crossing, and is well-described by a log-Gaussian, $P(\rm{capture}|\mathscr{B}) = A \exp[-(\ln \mathscr{B}-\mu)^2/2\sigma^2]$, with $A = 0.41^{+0.04}_{-0.04}$, $\mu = 1.09^{+0.08}_{-0.07}$, $\sigma = 1.05^{+0.08}_{-0.07}$. This fit, while an upper limit, is useful for simplified population synthesis. Finally, we explore the mass ratio AGN luminosity parameter space and find that GW190814 may be formed in a low luminosity AGN of $L_{\rm AGN}\approx 10^{43.5}\ \rm erg\ s^{-1}$. A more systematic parameter space exploration and future population studies will further test our predictions.

\end{abstract}


\section{INTRODUCTION }
\label{sec:Introduction}

Over 200 confident detections of gravitational wave events produced by the mergers of black hole and neutron star binaries have now been reported by the LIGO-Virgo-KAGRA collaboration (LVK) in the Fourth Gravitational Wave Transient Catalog (GWTC-4, \citealp{abac2025gwtc}). This census of compact binary objects in our Universe allows us to test our predictions about their expected mass distribution. The detected population has grown increasingly and unexpectedly diverse as our sample size has increased \citep{GWTC4_pop}. Of particular interest are events exhibiting mass ratios $q = m_2/m_1 < 0.5$, greater than those of 99 per cent of detectable binary black holes (BBH) \citep{Fishbach2020}. The most extreme of these is GW190814, with component masses $m_1=23.2\,M_{\odot}$ and $m_2=2.59\,M_{\odot}$, corresponding to a mass ratio $q=0.112^{+0.008}_{-0.009}$ \citep{abbott2020gw190814}. This event -- which is considered most likely to be a BBH merger \citep{Tews2021} -- was identified during the third LVK observing run. Explaining the origin of the highest mass ratio systems like GW190814 remains an outstanding problem for our stellar evolution and compact binary formation theories.

Despite the many models that have been proposed to explain the formation of BBHs \citep{bambi2022handbook,escriva2024black}, none have been completely successful in comprehensively explaining the rates and observed properties of the population of BBHs detected by the LVK \citep{abbott2023population}. The different binary formation channels can be split into two broad categories, isolated, and dynamical. Within the isolated channels BBH form from the collapse of massive stars within a binary, after which the binary evolves as a closed system with minimal interaction with any external environment (see, e.g. \cite{mandel2022merging} and references therein). The dynamical channel instead suggests that the black holes (BHs) are formed individually prior to the formation of the bound BBH, and evolve due to external perturbations, including a third object \citep[e.g.,][]{zwart1999black, sl17, ll19, tra2021,tra2022, man22, tra2024, liu2024, steg25, vg25,ginat2025}, repeating close encounters in dense environments such as globular \citep[e.g.,][]{rod18, sam18} and nuclear star clusters \citep[e.g.,][]{antonini12,gri18,hoang2018,tra2019,f19, knee2024, gri25lisa} and additional forces due to interaction with the ambient gas \citep[e.g.,][]{stone2017assisted, ginat2020, roz23, roz24, row23, rowan2024black, roz25}. 

A key prediction of the isolated channel is an upper (and sometimes also a lower) mass gap in the spectrum of BBH component masses in the range $\sim 50-130 M_\odot$ \citep{chatzopoulos2012effects, belczynski2016effect,ziegler2021filling, woosley2021pair, farag2022resolving,tani2021,tani2022}. This is in tension with data from the third observing run GWTC3 \citep[GWTC-3,][]{abbott2023population}, and contradicts observations of mergers of intermediate mass black holes like GW190521 with component masses $85^{+21}_{-14} M_{\odot}$ and $66^{+17}_{-18} M_{\odot}$ \citep{abbott2020gw190521}, or GW231123 with component masses $137^{+20}_{-19} M_{\odot}$ and $103^{+20}_{-52} M_{\odot}$ \citep{ligo2025gw231123}. In addition, this channel struggles to highly unequal mass ratio events \citep{safarzadeh2020being}, such as GW190814.

The dynamical channel, on the other hand, can explain both high mass and highly unequal mass ratio mergers. In particular, black hole mergers within the accretion discs of Active Galactic Nuclei (AGN) could allow for hierarchical mergers up to very large masses \citep{tagawa2020formation, rose2022formation, ford2022binary, atallah2023growing}. Other dynamical channels such as globular clusters struggle to retain remnant BHs in order to form high generational mergers \citep{rodriguez2019black, mapelli2021hierarchical, rozner2022binary}. In addition, globular clusters are unlikely to form highly unequal mergers \citep{gerosa2020astrophysical}.

It has been suggested that GW190814 could have been facilitated by an AGN disc \citep{abbott2020gw190814,yang2020black,mckernan2020black}. Many recent works focused on estimating the rates and properties of mergers in AGN discs \citep[e.g.,][]{mckernan2024mcfacts, vaccaro2025role}. However, these studies relied on simplified models of BH-BH and gas-BH interactions, neglecting many physical processes which are instrumental for of such evolution.

Gravitational interactions between the stellar mass BHs embedded in the AGN disc and the gas lead to a net torque, and consequently a net migration. The magnitude and direction of the migration sensitively depends on the disc and stellar mass BH (sBH) parameters. \cite{bellovary2016migration} showed that there can be locations in the disc where the net torque changes sign, leading to a `migration trap' where BHs can accumulate. This leads to a natural environment for BHs to pair up and merge rapidly, with no obvious mechanism preventing highly unequal mass ratio binary formation. \cite{grishin2024effect} showed that the inclusion of thermal torques significantly influence where migration traps form, or indeed if they form at all. In addition, \cite{gilbaum2025escape} showed that the gap opening of more massive black holes in the AGN disc also strongly affects which sBH have migration traps in a given AGN disc. Together, these call into question the feasibility of producing highly unequal mass ratio mergers in AGN discs. However, \cite{gilbaum2025escape} relied on semi-analytical prescriptions of the timescales for realignment and migration in the AGN disc and BBH mergers in migration traps, largely ignoring the non-linear and often chaotic dynamics of BH migration and binary capture in AGN discs. In addition, they did not account for how the formation of mean motion and co-orbital resonances can prevent the merger of black holes during encounters at traps \citep{secunda19, epstein-mmr2025}. All of these dynamical interactions are important to consider for the origin of GW190814 and other extreme events, as they can substantially affect the rate of AGN-disc mediated mergers.

In this paper, we investigate the dynamics of highly unequal mass ratio BBH mergers in AGN discs by coupling AGN disc torques with an N-body code. This gives a more complete view of the coupled N-body and gas dynamics in AGN discs, particularly focusing on the different dynamical outcomes that can prevent mergers depending on the specific AGN disc and radial location. To do this, we use the public \texttt{pAGN} package \cite{pAGN-code} to generate the AGN model and torque prescription, and couple these forces with the N-body code \texttt{TSUNAMI} \citep{tsunami-code}. We use this code to explore the putative origin of GW190814 as a black hole binary formed via interactions with an AGN disc.

Our paper is organised as follows: in Sec. \ref{sec:Background} we overview the underlying physics and parameter space of AGN discs. In Sec. \ref{sec:Methods} we present the computational techniques. In Sec. \ref{sec:Results} we show the different dynamical outcomes of encounters of highly unequal mass BHs in an AGN disc. Sec. \ref{sec:analytic_presc} is devoted to developing a semi-analytic model and empirical fitting of the capture probability based on a comprehensive population study of integrating many systems. In Sec. \ref{sec:discussion} we discuss our limitations and future work, and finally summarise our main key outputs in Sec. \ref{sec:conclusion}.

\section{BACKGROUND THEORY}
\label{sec:Background}

\subsection{Behavior of traps across AGN discs}

A crucial element of AGN discs is the existence and location of migration traps, where most of BH mergers are believed to occur. These are locations in the AGN disc where the torque on an embedded BH changes sign from negative (inward migration) to positive (outward migration), leading to a stable equilibrium \citep{bellovary2016migration}. There can additionally be unstable equilibrium points, called `anti-traps', where the torque changes sign from positive to negative. BHs are expected to `park' at the migration traps and accumulate to a local over-density which efficiently leads to rapid pair-ups and mergers. Earlier models suggest that the location of these traps is independent of the perturber (i.e. black hole) mass, and mostly depend on the `density bump', which occurs near the transition from radiation dominated pressure in the hot inner zones to gas dominated pressure in the cooler zones (see, e.g., \citealp{2003MNRAS.341..501S} for details).

Recent work by \cite{grishin2024effect} accounted for thermal effects on the total torques (using the prescription of \citealp{jimenez2017improved}), which significantly alter the existence and location of migration traps. In addition, massive black holes can open up gaps in the AGN disc, which \cite{gilbaum2025escape} showed strongly affects which BHs are trapped. This lead the authors to conclude that hierarchical mergers can only form from traps if the AGN is within a certain luminosity range, provided that mergers occur primarily at trap locations.

In order to explore the details of BH migration in AGN discs we need to identify representative AGN disc models. Let $m_1$ and $m_2$ be two different BH masses ($m_1>m_2$). Here we focus on the highly unequal mass ratio event of GW190814, but the features are generic and a more detailed population study will be carried out in the future. The initial conditions for the AGN disc is the SMBH mass $M_{\rm SMBH}$, the accretion rate in Eddington units $\dot{m} =\dot{M}_{\rm SMBH}/\dot{M}_{\rm Edd}$, and the \cite{shakura1973black} $\alpha$ viscosity parameter. For AGN with high luminosity, migration traps don't exist at all and all BH are expected to eventually inspiral to the SMBH and create a wet EMRI. Otherwise, the existence and location of the trap can depend on the mass of the sBH: Above a critical mass $m_{\rm gap}$, the gap formed in the AGN disc will lead to the elimination of the thermal effects and the BH will strictly migrate inward. The gap mass is given by \cite{kanagawa2018radial}
\begin{equation}\label{eq:gap_opening_mass}
    m_{\rm{gap}} = 25 M_{\rm SMBH}\sqrt{\alpha (H/R)^5},
\end{equation}
where $H/R$ is the local disc aspect ratio. The behaviour will be qualitatively different depending on the ordering of the masses: If $m_{\rm gap} > m_1$, both masses are expected to be trapped. If $m_{\rm gap} < m_2$, neither of the masses is expected to be trapped and both will inspiral to the SMBH. Finally, if $m_1>m_{\rm gap}>m_2$, only the least massive BH will be trapped and a close encounter is expected when $m_1$ approaches the $m_2$ trap.

\begin{figure}
    \includegraphics[width=0.50\textwidth]{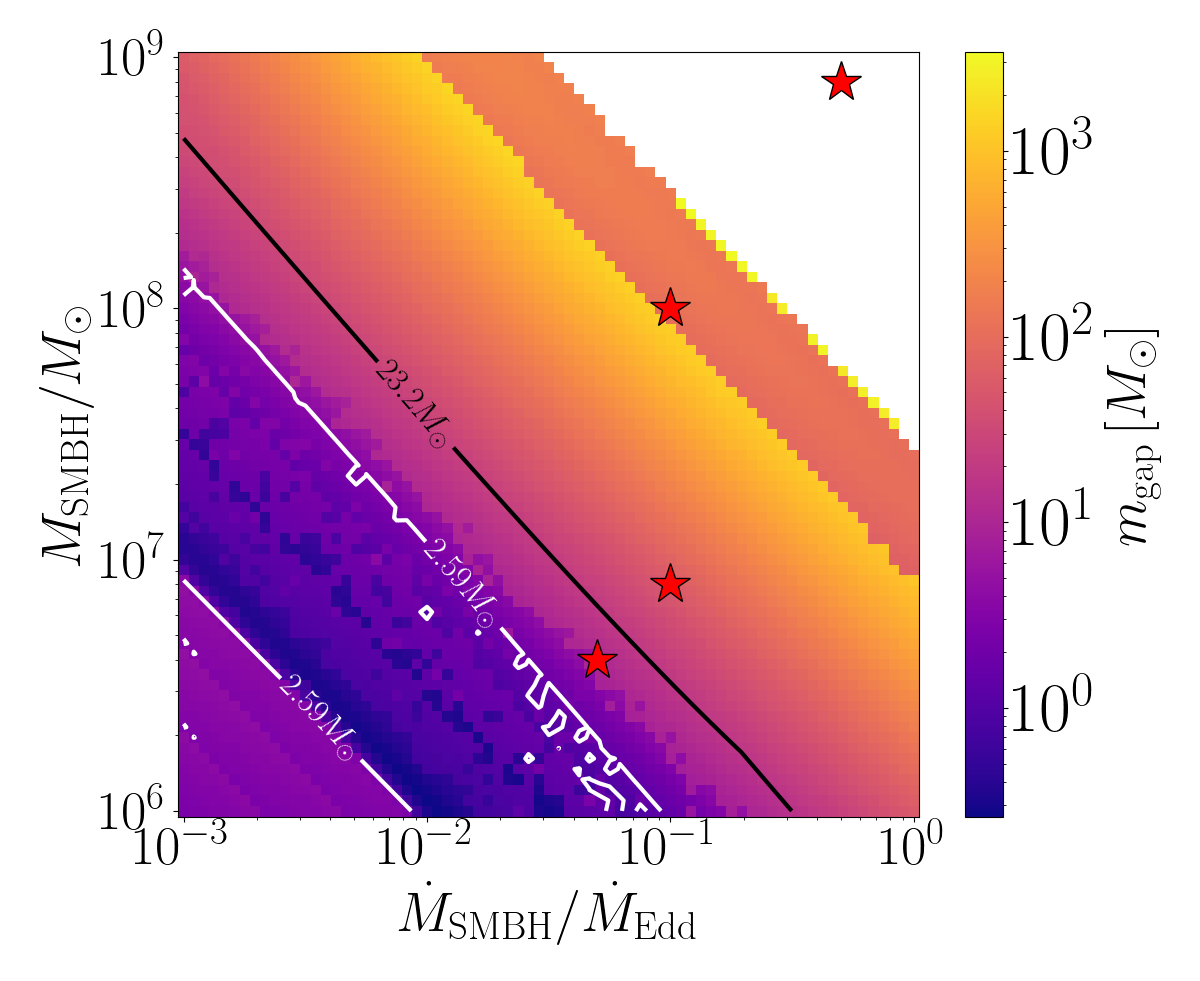}
    \caption{Mass gap as a function of accretion rate and SMBH mass, for a fixed viscosity parameter $\alpha=0.01$. The heat map instensity indicates the minimum gap opening mass, $m_{\rm gap}$, computed for each disc by evaluating Eqn. \ref{eq:gap_opening_mass} at the location between the outer migration trap and anti-trap. The contour lines shows the parameter space where $m_{\rm gap}$ matches the masses of interest, $m_{\rm gap}=2.59M_{\odot}$ or $m_{\rm gap}=23.2M_{\odot}$, giving a proxy for the discs that will produce migration traps for these masses. The stars represent the parameter values of the discs we consider in the study.}
    \label{fig:Mgap_hr}
\end{figure}

In order to explore the properties of migrating BHs for representative AGN discs, we first map out the parameter space. We identify four types of representative AGN discs depending on $m_{\rm gap}$ (if traps exist). A systematic study of the AGN parameter space is deferred for a future study. For each disc, we use \texttt{pAGN} to solve for the disc profile, allowing us to compute the location of all migration traps and anti-traps. We then use Eqn. \ref{eq:gap_opening_mass} to compute the maximum gap opening mass, $m_{\rm gap}$, of the disc in the region of positive torque (i.e., between the migration trap and anti-trap). We show this $m_{\rm gap}$ for each disc in the parameter space as a heat map in Figure \ref{fig:Mgap_hr}. This is similar to Figure 3 of \cite{gilbaum2025escape}, with the colour intensity indicating the gap opening mass for the given disc parameters.

We overlay the heat map with contours of $m_{\rm gap}=2.59M_{\odot}$ and $m_{\rm gap} = 23.2 M_{\odot}$, which partitions the parameter space into discs where neither of the GW190814 masses contain traps, only one object is trapped ($2.59 M_{\odot} \leq m_{\rm gap} \leq 23.2 M_{\odot}$), or both objects are trapped ($m_{\rm gap} \geq 23.2 M_{\odot}$). Note that the uncoloured region in the top right of Figure \ref{fig:Mgap_hr} shows discs where no traps exist, with thermal torques being subdominant and hence migration remains inward throughout the whole disc \citep{grishin2024effect}. We additionally note that the $m_{\rm gap}$ contours approximately partition the $M_{\rm SMBH}-\dot{M}_{\rm SMBH}$ parameter space with diagonal lines of constant $M_{\rm SMBH} \cdot \dot{M}_{\rm SMBH}$ - which is proportional to the AGN luminosity:

\begin{align}
    L_{\rm AGN} = \eta \dot{M}_{\rm SMBH} c^2 
    \approx 10^{44}\left( \frac{\eta}{0.1} \right) \left( \frac{\dot{m}}{0.1} \right) \left( \frac{M_{\rm SMBH}}{7\times 10^6 M_\odot} \right)\ \rm erg\ s^{-1},
    \label{eq:L_AGN}
\end{align}
\noindent
where we assume the radiative efficiency is $\eta = 0.1$ for a non-rotating SMBH. The dependence on $M_{\rm SMBH}$ comes from rescaling the accretion rate to Eddington units \citep[see e.g.,][]{gilbaum2022feedback, grishin2024effect}. 

\begin{figure*}
    \centering
    \includegraphics[width=\textwidth]{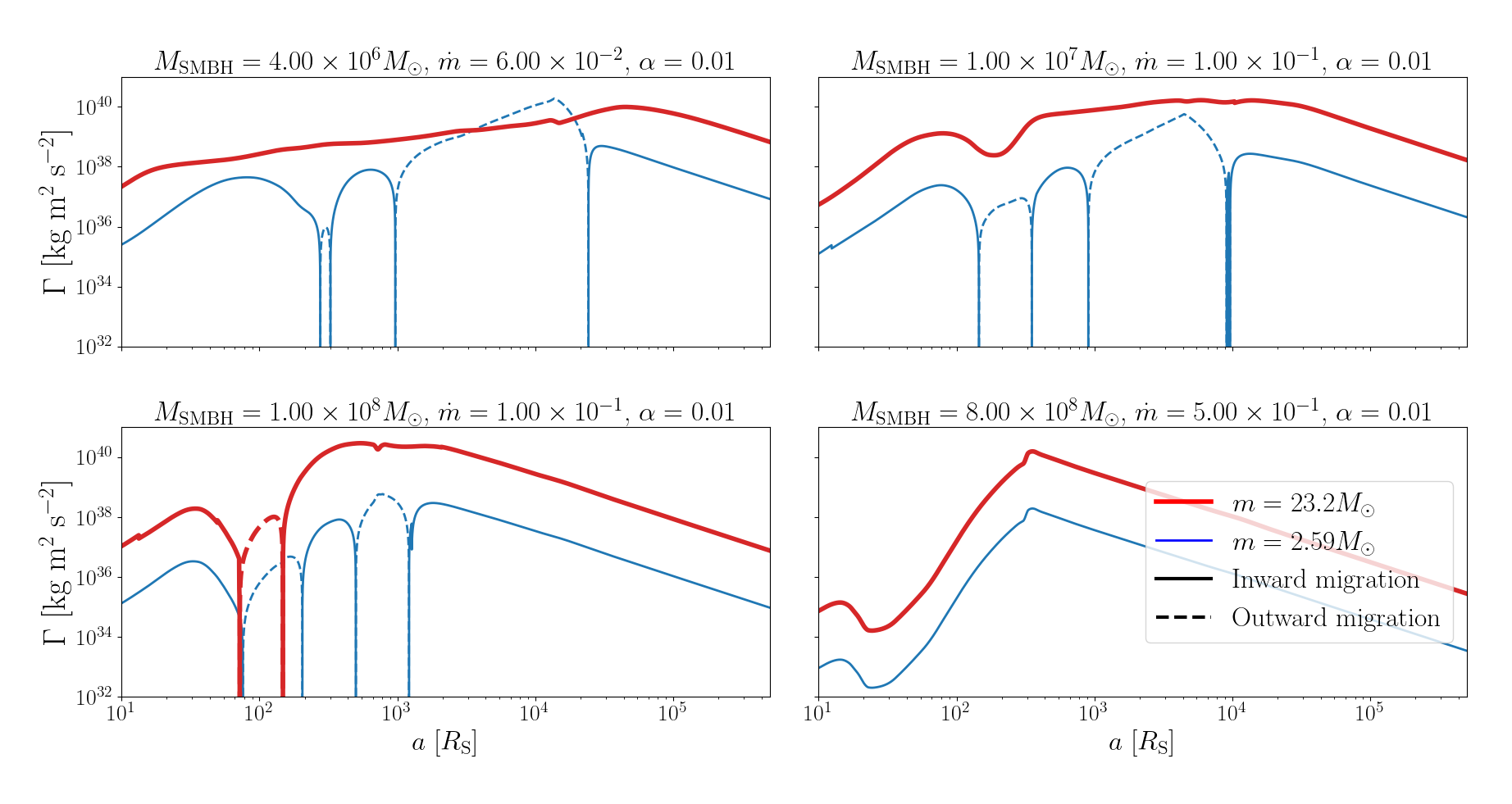}
    \caption{The absolute value of the migration torques for $m_1$ and $m_2$ for each disc of interest. These torques include the effects of gap opening and thermal torques, shown in Eqn. \ref{eq:torque_tot}, with full expressions given in the appendix of \protect\cite{gilbaum2025escape}. The regions of negative torque (resulting in inward migration) are shown with solid lines, while the regions of positive torque (leading to inward migration) are shown as dashed lines.}
    \label{fig:type_2_torques_plots}
\end{figure*}

Figure \ref{fig:type_2_torques_plots} shows the radial torque structure of the four representative AGN discs. The top left panel shows the  disc with $m_2 < m_{\rm gap} < m_1$ such that only the lower mass $m_2$ has a migration trap. The bottom left panel shows the disc model with $m_{\rm gap} > m_1$ so both BH have migration traps. The top right panels shows a similar situation to the top left one, even though formally $m_{\rm gap} > m_2$. This is due to the fact even a partial gap may change the delicate balance between the gravitational (type I and type II) and thermal torques, and the estimate of $m_{\rm gap}$ without the thermal effects is inaccurate. Finally, the bottom right panel shows the disc that has no migration traps.

\subsection{Mean motion resonances}

$(k+p):k$ mean motion resonances (MMR) arise when the orbital periods of two orbiting bodies form an integer ratio, 
\begin{equation}\label{eq:resonance_k}
    \frac{P_1}{P_2} = \frac{k+p}{k}, \quad k,p \in \mathbb{N},
\end{equation}
which gives rise to a slowly varying resonant angle, 
\begin{equation}
    \phi = (k+p)\lambda_2 - k\lambda_1 ,
\end{equation}
where $\lambda_i$ are the mean longitudes of the two bodies, and $p$ is the order of the resonance. Such resonances are central to planetary dynamics, as they can maintain stable orbital configurations over long timescales. A well-known example is the $3{:}2$ resonance between Pluto and Neptune, which prevents close encounters despite their intersecting orbits \citep[e.g.,][]{malhotra95}. In protoplanetary discs, MMRs are thought to have shaped the final orbital architecture of many planetary systems \citep{murray1999solar,armitage2020astrophysics}. Thus, in this context, there have been extensive studies on the conditions in which convergent migration leads to resonance capture
\citep{tremaine2023dynamics,lin2025resonance}.

The migration of BHs in AGN discs is closely analogous to planetary migration in protoplanetary discs \citep{baruteau2013recent}. Thus, MMRs are expected to form when two BHs migrate on converging orbits. Despite this, population studies of BBH in AGN discs have so far largely relied on simplified one-dimensional dynamical prescriptions that neglect the possible influence of resonances \citep{mckernan2024mcfacts,vaccaro2025role}. However, resonant interactions may play an important role in determining encounter outcomes and ultimately the merger rates of BH in AGN environments \citep{secunda19,lott2025scattering,epstein-mmr2025}.

\section{COMPUTATIONAL METHODS}
\label{sec:Methods}

\subsection{\texttt{TSUNAMI} N-body code}

The simulations in this work are based on  \texttt{TSUNAMI}, an N-body code that efficiently solves the gravitational equations of motion with a high level of precision \citep{tsunami-code}. The code minimizes numerical round-off errors through regularization of the equations of motion (eliminating the singularity arising as the distance between bodies goes to zero), chain coordinates (useful for hierarchical systems, where close by particles are far from the center of mass), and Bulirsch–Stoer extrapolation (allowing for accurate integration over wide range of time scales). \texttt{TSUNAMI} has additional post-Newtonian corrections up to 3.5th order \citep[3.5PN,][]{,blanchet2014gravitational}, necessary for the relativistic dynamics of binary black holes, and nearby the central supermassive black hole in the AGN disc case.

\subsection{AGN disc model and torque prescription}

In this paper, we include external torques to the bodies due to gravitational interactions with the gas present in AGN accretion discs. We use the AGN disc model of \cite{2003MNRAS.341..501S}, as implemented in the \texttt{pAGN} package \cite{pAGN-code}. This model parametrizes the properties of AGN discs (such as temperature, density, scale height, etc) as a function of distance from the central SMBH. It comprises a classic geometrically thin, optically thick accretion discs of \cite{shakura1973black}, with the additional assumption of some heating source present in the disc to marginally support the outer regions from gravitational collapse.

Given the disc properties at the bodies location, the torques are computed using the same prescription described in \cite{gilbaum2025escape}. This includes the unified type I and type II torques described in \cite{kanagawa2018radial}, in addition to thermal torques \citep{masset2017coorbital}.

\subsection{Coupling the codes}\label{sec:coupling_codes}

In our simulations, for a given mass $m$ the additional, mass dependent torques $\Gamma_{\rm tot}(m)$, computed from the AGN disc model lead to migration on a time scale:

\begin{equation}
    \tau_{\rm{mig{}}} = \frac{-L_{\rm tot}}{2 \Gamma_{\rm{tot}}} \exp \left( \frac{z^2}{2 H^2} \right),
\end{equation}

\noindent
where $L_{\rm tot}$ is the orbital angular momentum of the BH, $H$ is the local disc height, and $z$ is the height of the particle above the disc midplane. 
\noindent
The total torque is computed via
\begin{equation}\label{eq:torque_tot}
    \Gamma_{\rm{tot}}=\Gamma_{\rm II} + \exp(-K/25) \Gamma'_{\rm{th}},
\end{equation}
where the expressions for the thermal torque, $\Gamma'_{\rm th}$, and reduced surface density type II torque:
\begin{equation}\label{eq:type_II_torque}
    \Gamma_{\rm II} = \frac{\Gamma_{\rm L}+\exp(-K/20) \Gamma_{\rm C}}{1+K/25},
\end{equation}
\noindent
can be found in appendix section A of \citep{gilbaum2025escape}, while $K=25 (m/m_{\rm gap})^2$.

In addition to the radial torques, we implement eccentricity and inclination damping timescales  \citep{cresswell2006evolution,kanagawa2020radial}:

\begin{equation}
    \tau_e=\frac{\tau_{\rm{mig}}}{0.780} \left(  \frac{H}{R}  \right)^2,
\end{equation}

\begin{equation}
    \tau_\iota=\frac{\tau_{\rm{mig}}}{0.540} \left(  \frac{H}{R}  \right)^2.
\end{equation}

\noindent
This leads to an evolution of the orbital parameters:

\begin{equation}
    \dot{a} = \frac{-2 a}{\tau_{\rm mig}},
\end{equation}

\noindent
and similarly for $e$ and $\iota$. The corresponding additional velocity dependent acceleration is \citep{kajtazi2023mean}:

\begin{equation}
    \dot{\bf{a}} = -\frac{\bf{v}}{\tau_{\rm mig}}-\frac{2 (\bf{v} \cdot \bf{r}) \bf{r}}{r^2 \tau_e}-\frac{(\bf{v} \cdot \hat{z})}{\tau_\iota} \bf{\hat{z}},
\end{equation}
where $\bf{r}$ and $\bf{v}$ are the position and velocity vector in the cylindrical reference frame aligned with the total angular momentum of the AGN disc. This is akin to a gas dynamical friction effect that tends to decrease the relative velocity, and is especially strong once $e,\iota$ are comparable to the disc aspect ratio \citep{ostriker99,muto11,  gri15, samsing_gdf}.

The simulations are initialized with a central SMBH, stellar BH with masses $m_1=23.2 M_{\odot}$ and $m_2=2.59M_{\odot}$ initialised on Keplerian orbits about the central SMBH. The initial {eccentricities and inclinations} are sampled from Rayleigh distributions, $x=\{e_{i}, \iota_i\} \in {\rm Rayleigh}(x, \sigma)$, where the standard deviation $\sigma = H/R$ is the local aspect ratio, and the probability density function is $P(x) = {\rm Rayleigh}(x,\sigma)=x/\sigma^2 \exp(-x^2/2\sigma^2)$. The other orbital parameters are uniformly randomly distributed; $\omega_i, \Omega_i, \nu_i \in {\rm Uniform}(0, 2\pi)$.

\section{ORBITAL ENCOUNTER DYNAMICS}
\label{sec:Results}

\subsection{Orbital encounter scenarios}
In this section we describe the possible outcomes when masses $m_1$ and $m_2$ migrating through the disc encounter one another. We observe each of the following end states:

\begin{enumerate}
    \item Binary formation and merger,
    \item Orbit crossing,
    \item Mean Motion Resonance capture.
\end{enumerate}
\noindent
Another transient co-orbital 1:1 `tadpole' resonances may form. As the BBH continues to migrate it will eventually be broken into either (ii) Orbit crossing or (i) Binary formation and merger end states.  Contrary to the intuition of 1D discs, a close encounter in the disc crossing does not necessarily lead to capture and merger (and neither does resonance capture). In this section, we describe the qualitative dynamics and individual examples for the different outcomes. A more detailed population study and semi-analytical modelling of the statistical outcomes are presented in in Sec. \ref{sec:analytic_presc}.

\subsection{Binary capture and merger}

\begin{figure}
    \centering
    \includegraphics[width=0.5\textwidth]{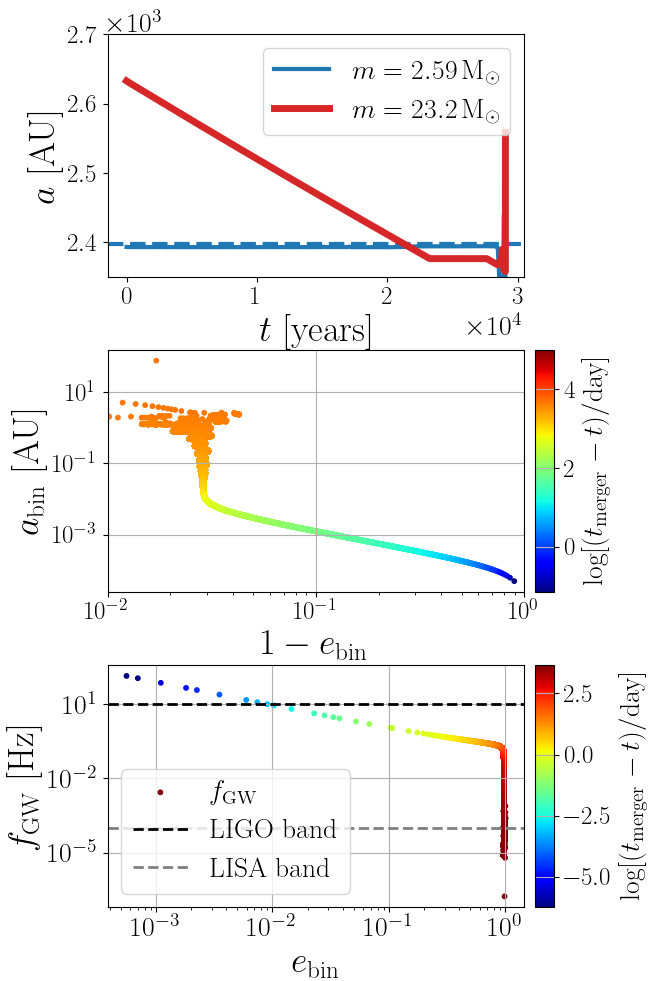}
    \caption{Top: orbital evolution of $m_1 = 2.59M_{\odot}$ (blue) and $m_2 = 23.2 M_{\odot}$ (red). Dashed lines are trap locations. Middle: semi-major axis and eccentricity of the inner binary from capture to merger, colour-coded with the time before merger. Bottom: The evolution of the peak gravitational wave frequency $f_{\rm GW}$ with eccentricity.}
    \label{fig:orb_evo_merger}
\end{figure}

We show a representative example of a binary capture and merger, by initialising a $m_2=2.59 M_{\odot}$ BH at its trap location of distance $r_{2,\rm trap} \approx 9\times10^3\  \rm{AU}$ from a central SMBH of mass $M_{\rm SMBH}=10^8 M_{\odot}$ and accretion rate of $\dot{m}=0.1$ in Eddington units. We set the viscosity parameter $\alpha=0.01$ throughout this work. We initialise a BH of mass $m_1=23.2 M_{\odot}$ at $r \approx 2.6 \times 10^3\  \rm{AU}$, which begins migrating towards its trap, located at $r_{1,\rm trap} \approx 2.4\times 10^3\  \rm{AU}$, due to migration torques. The subsequent orbital evolution of a single case is shown in the top panel of Figure \ref{fig:orb_evo_merger}, where we observe a binary being formed around $t=3 \times 10^4$ years after the beginning of the simulation, and a subsequent merger of $m_1$ and $m_2$ on a much shorter timescale.

A highly eccentric binary is formed, leading to a rapid binary inspiral that lasts $\approx 44$ years, corresponding to approximately 10 orbits about the central SMBH. The evolution of the semi-major axis and eccentricity of the $m_1-m_2$ binary is shown at the bottom of Figure \ref{fig:orb_evo_merger}. The initial binary evolution is chaotic, showing rapid oscillations in the eccentricity during the initial gas-dominated inspiral regime, until the eccentricity stabilizes at a value of $e_{\rm bin} \sim 0.97$, followed by GW-dominated inspiral. During this regime, the binary eccentricity is greatly dampened from gravitational wave emission, with $e_{\rm bin}=0.1$ when the orbital separation $d < 20 r_{\rm g}$, where $r_{\rm g}=G(m_1+m_2)/c^2$.

To calculate the eccentricity of the binary when the GW signal enters the audio band ($10-2000\,\mathrm{Hz}$), we take orbital parameters of the binary from TSUNAMI when $d=20 r_{\rm g}$, and evolve the system to the inspiral regime by solving the \cite{Peters64} equations for $(a,e,\omega)$ evolution under GW emission. Then, using \cite{wen2003eccentricity} we can estimate the peak frequency of GW emission $f_{\rm GW}$. Plotting the resulting evolution of $f_{\rm GW}$ with $e_{\rm bin}$ in the bottom panel of Figure \ref{fig:orb_evo_merger}, we see that $e_{\rm bin} \approx 0.01$ at $f_{\rm GW}=10 \rm{Hz}$.

\subsection{Orbit crossing and resonance traps} \label{sec:passby}

\begin{figure*}
    \includegraphics[width=\textwidth]{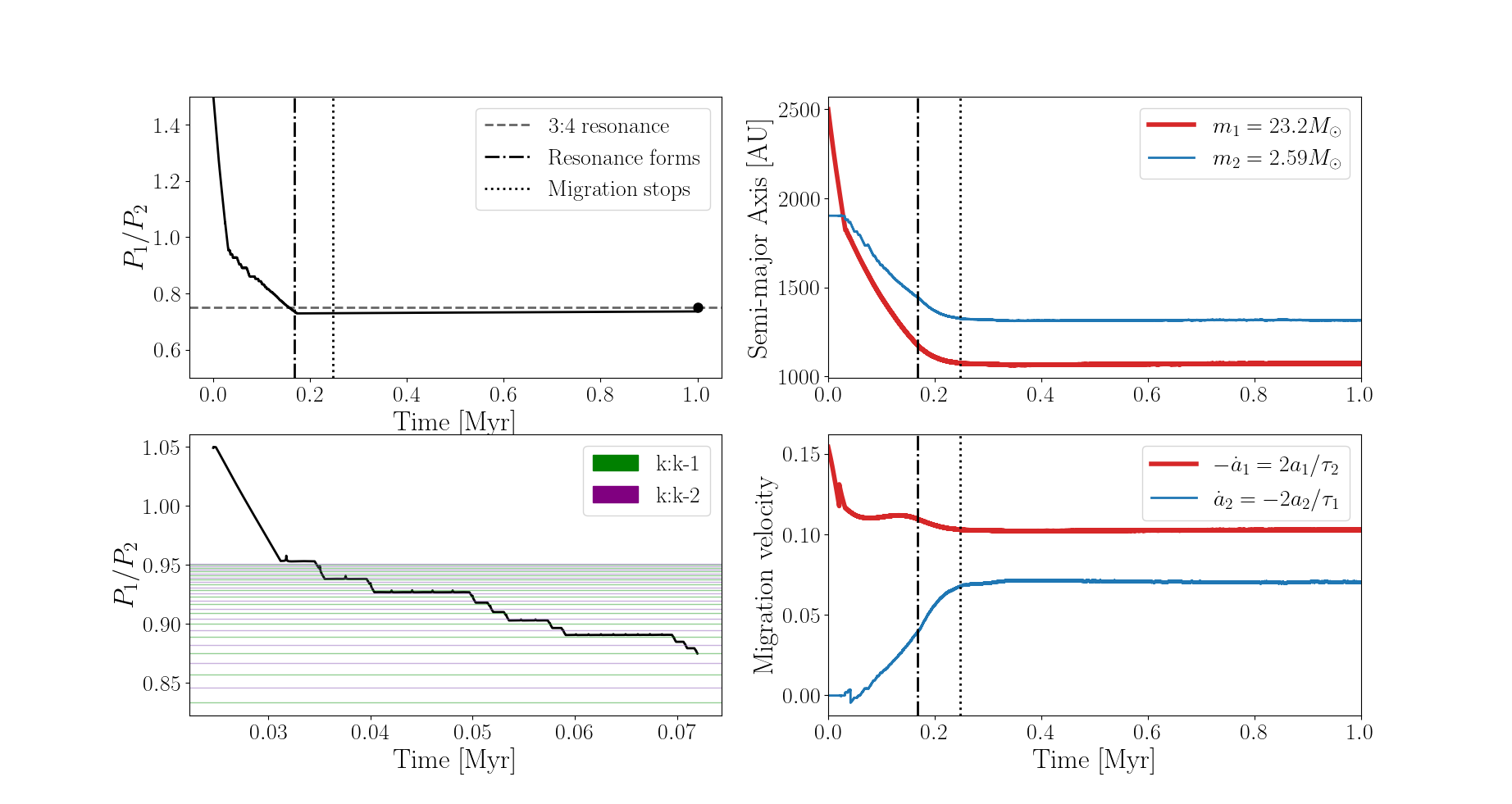}
    \caption{An example of an orbit crossing, during an encounter at a trap in the disc $(M_{\rm SMBH}/ M_{\odot},\dot{m}, \alpha)=(10^8, 0.1, 0.01)$. In this scenario, long after an orbit crossing, a 3:4 resonance forms, synchronizing the inspiral of the two bodies.}
    \label{fig:resonance_trap_orbit}
\end{figure*}

When the massive BH ($m_1$) migrates fast compared to the migration of $m_2$, orbit crossing is most likely to occur. Figure \ref{fig:resonance_trap_orbit} shows an example where the massive BHs orbit crosses over the secondary BH. In this case, divergent migration then leads the system into capture into a MMR where the period ratio stabilizes to $4:3$ (top left). During the inspiral phase after the orbit crossing, the system is temporarily captured into a cascade of first and second order MMRs (bottom left). These resonances are not stable, most likely due to the changing relative migration torques as the black holes migrate through the disc. The semi-major axis evolution shows that the migration effectively halts (top right). In this case, the torques from the gas are counter-balanced by the resonant interactions (top right), leading to a stable configuration.

\begin{figure}
    \centering
    \includegraphics[width=0.5\textwidth]{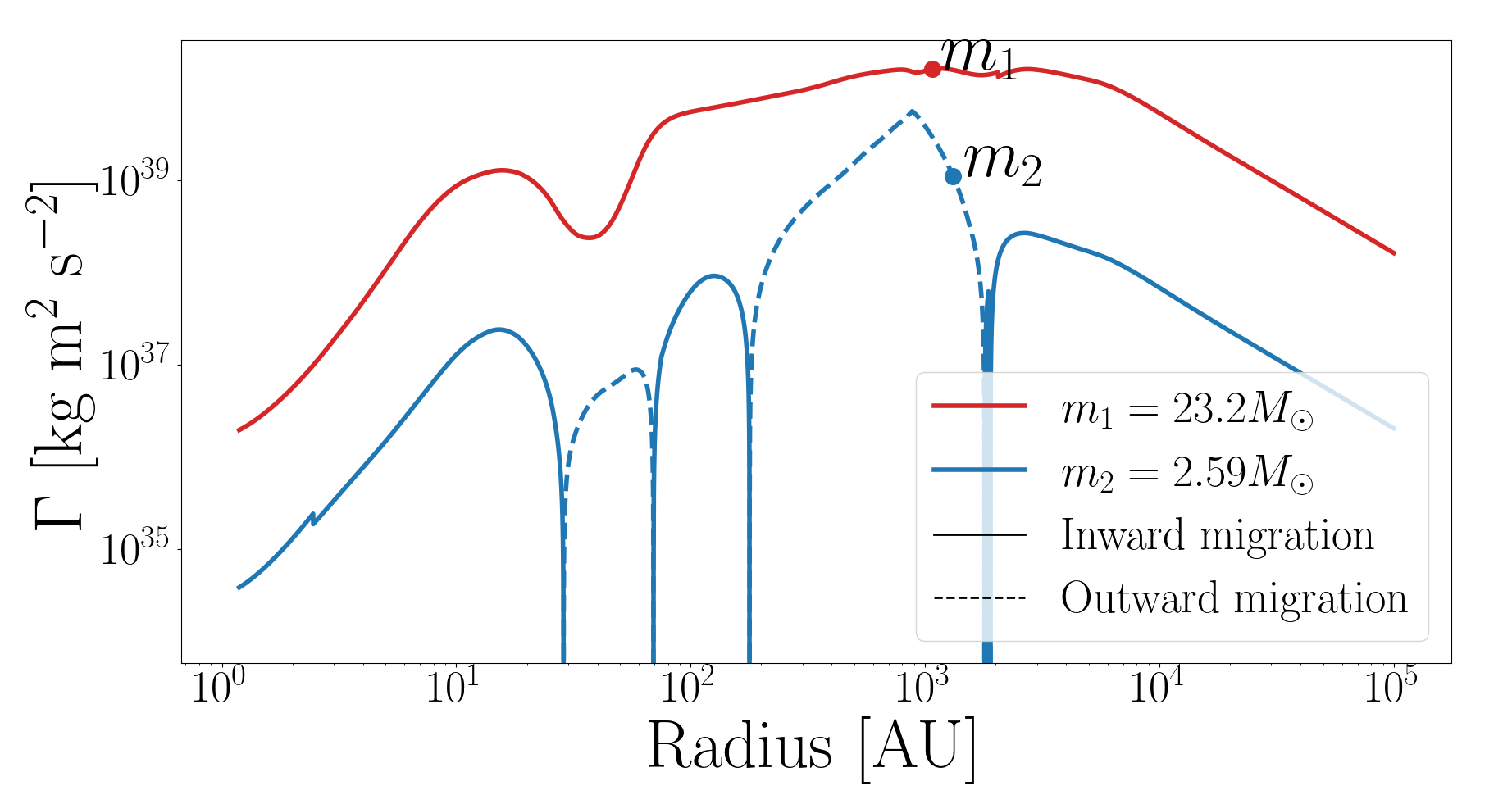}
    \caption{Resonance trap final locations (see Figure \ref{fig:resonance_trap_orbit}). We observe that the final radial locations of each BH does not coincide with any trap location, with the stability of the orbits being maintained by the resonant interaction and the opposing migrations from the gas torques.}
    \label{fig:resonance_trap_torque}
\end{figure}

In this remarkable scenario, the more massive $m_1$ is held in place by the resonant interaction with a mass about $1/10$-th of $m_1$. The gas torques causing the migration of each BH are non-zero -- the gas torques on $m_2$ are negative, while the torques on $m_1$ are positive (Figure \ref{fig:resonance_trap_torque}) -- yet the two objects are essentially trapped due to the resonant interaction.

\begin{figure*}
    \centering
    \includegraphics[width=\textwidth]{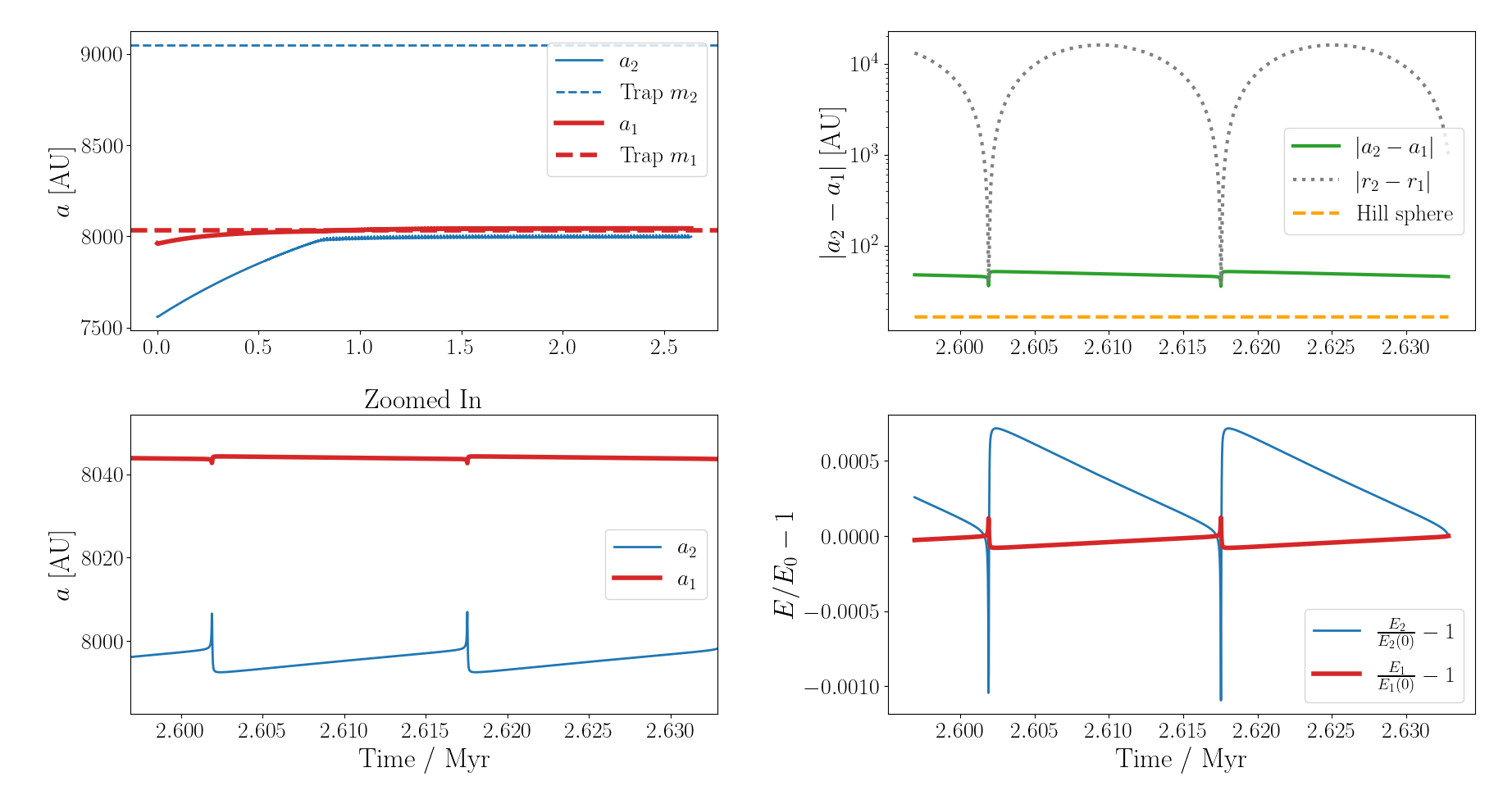}
    \caption{Illustration of Hill limit resonance capture near a trap. Notice that the resonance is created due to asymmetric energy exchange between $m_1$ and $m_2$ during conjunction. This prevents the orbits of the two black holes from every crossing. There is a balance between the energy exchange during conjunction with the energy change between conjunctions (due to gas torques). This leads to a long term stable resonance that prevents a merger.}
    \label{fig:resonance_close_enc}
\end{figure*}

\subsection{Resonance blocking in traps}

We now consider the opposite limit, where the relative migration rate at the trap location is small. Despite the proximity of the two BHs, high $k$ (Eqn. \ref{eq:resonance_k}) first order resonance can still prevent a close encounter.
To illustrate this, we show a simulation in the disc $(M_{\rm SMBH}/M_{\odot}, \dot{m}, \alpha)=(10^7, 0.1, 0.01)$, with initial conditions such that the migration is outward. The result is that $m_2$ is prevented from crossing the $m_1$ migration trap, due to the formation of a mean motion resonance (MMR), as shown in Figure  \ref{fig:resonance_close_enc}.

\begin{figure*}
    \centering
    \includegraphics[width=1.05\textwidth]{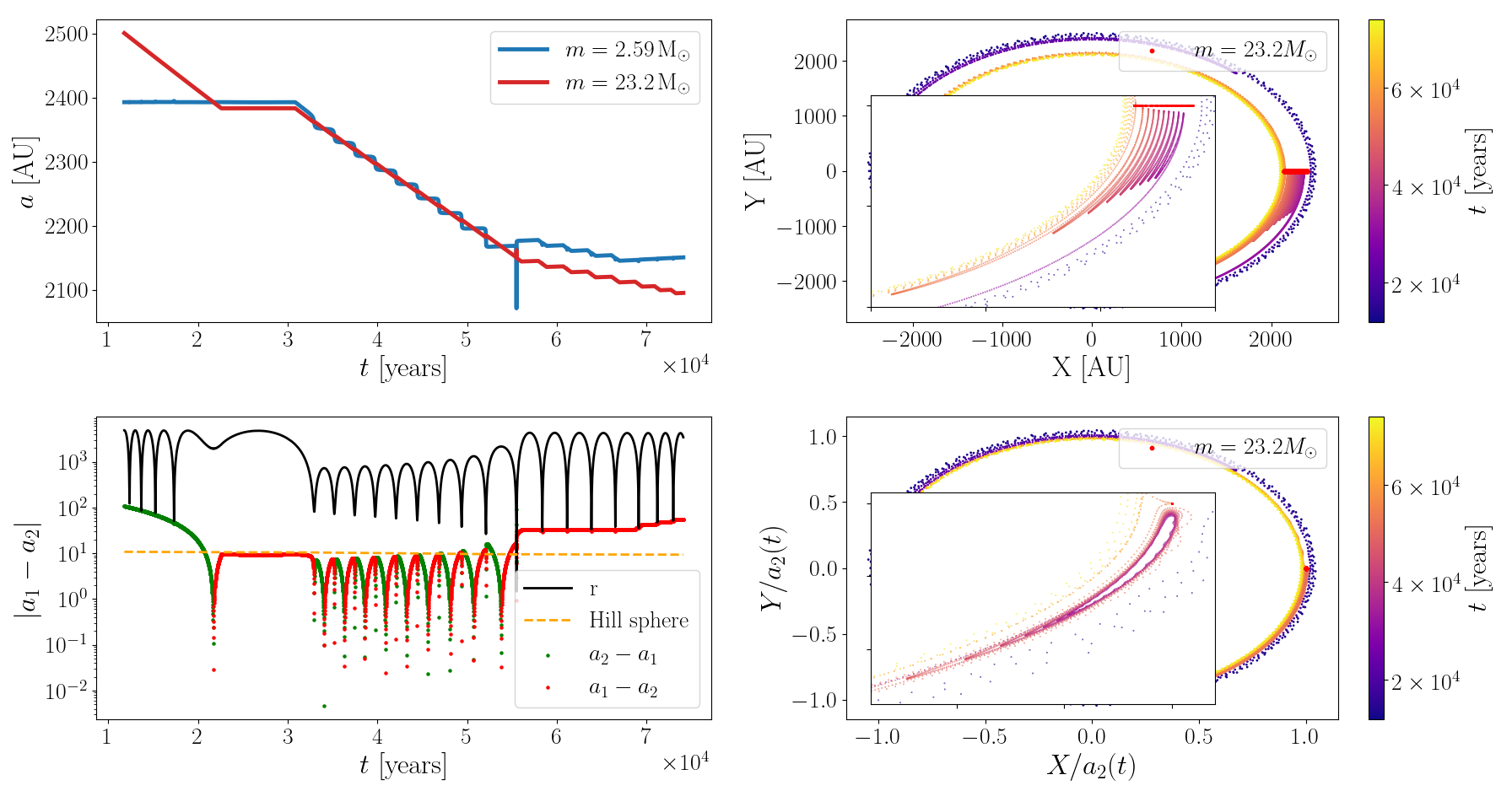}
    \caption{An example of an encounter that results in a tadpole orbit forming. The top left shows the evolution of each of the semi-major axis. The bottom left figure shows the orbital separation $|a_1-a_2|$ between the orbits as a scatter plot, compared with the solid line $r$ showing the spatial distance between the bodies with time. The right two figures show the spatial evolution of the mass $m_2=2.59M_{\odot}$ librating in a tadpole about $m_1=23.2 M_{\odot}$. Both right figures are shown in the co-rotating frame of $m_1$, but the bottom left is additionally normalised by the semi-major axis of $m_1$.}
    \label{fig:tad_pass}
\end{figure*}

\begin{figure*}
    \centering
    \includegraphics[width=\textwidth]{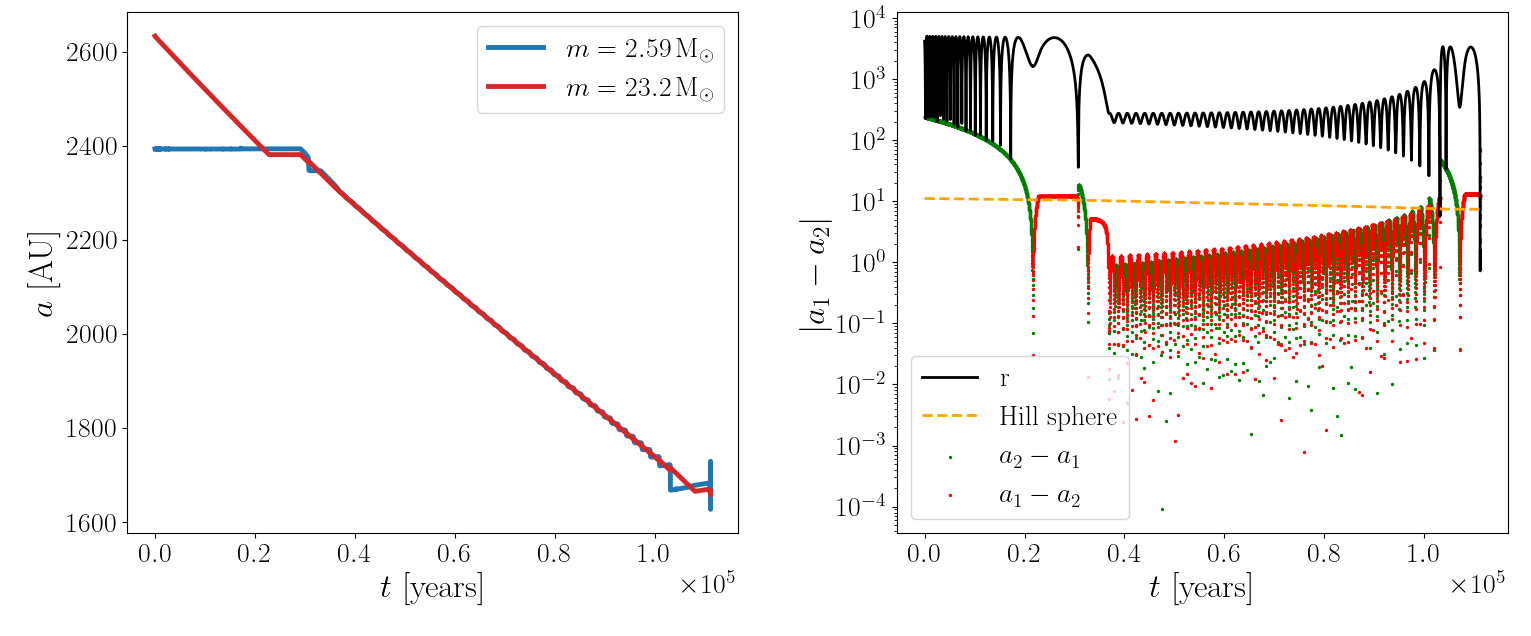}
    \caption{The evolution of an orbit that begins as a tadpole orbit, but becomes unstable after some time. After the co-orbital resonance is broken, a binary is formed, and subsequently there is a binary capture and merger at a location well separated from the encounter location.}
    \label{fig:tadpole_to_merger}
\end{figure*}

Unlike the $4:3$ MMR seen in Figure \ref{fig:resonance_trap_orbit}, which had a period ratio $P_1/P_2 \approx 0.75$, the period ratio of the two orbits is $P_2/P_1 \approx k/(k+1)$ for $k\approx100$\footnote{Finding a librating resonant angle with external torques is difficult when the orbital migration is significant on the timescale of one orbit \citep{petit2020resonance}.}. Thus, during orbital conjunctions, the distances between the binaries can temporarily approach the Hill radius:
\begin{equation}
    R_{\rm H} = \frac{a_1+a_2}{2} \left( \frac{m_1+m_2}{3 M_{\rm SMBH}} \right)^{1/3},
\end{equation}
as can be seen in Figure \ref{fig:resonance_close_enc}.  
These high $k$ resonances can be explained simply in the Hill limit, as described in \cite{tamayo2025unified} in the context of planetary systems. During the brief period leading up to conjunction, the inner BH $m_2$ loses energy, $|E_2|=G M_{\rm SMBH}m_2/a_2$, to $m_1$ ($a_2$ increases), while $m_2$ gains energy after conjunction ($a_2$ decreases). The asymmetry in this energy exchange before and after conjunction leads to a net increase in energy of the $m_2$ orbit ($a_{1,\rm f}<a_{1,\rm i}$) and a net decrease in energy of the $m_1$ orbit (bottom right panel of Fig. \ref{fig:resonance_close_enc}). This net change in orbital energy's before and after conjunction is counteracted by energy increases of $m_1$ (from negative gas torques) and energy decrease in $m_2$ (from positive gas torques) between conjunctions. This leads to an equilibrium at the trap which is stable for the lifetime of the simulations, preventing the orbit of $m_2$ from crossing that of $m_1$. 

In other words, the there is a damping/repulsive cycle for each synodic period: The damping due to migration torques brings the orbits together and prevents eccentricity growth, while at conjunction there is a two-body impulse where the two black holes repulse each other and the cycle begins again.

\subsection{Co-orbital `tadpole' resonances}

If an orbital crossing is not blocked by a mean motion resonance, co-orbital resonances can also form, where the period ratio librates about $P_1/P_2=1$. An example of this is shown in Figure \ref{fig:tad_pass}, where a close encounter in the disc $(M_{\rm SMBH}/M_{\odot}, \dot{m},\alpha)=(10^8, 0,1, 0.01)$ results in $m_2$ forming a tadpole orbit around $m_1$, with the combined $m_1-m_2$ system migrating inward. 

As the tadpole system migrates, the changing gas torques lead to an evolving tadpole libration amplitude (Figure \ref{fig:tad_pass}), in agreement with analytic models \citep{sicardy2003co}. Eventually, the libration amplitude becomes too great, and the system becomes unstable, resulting in the tadpole breaking. After the tadpole breaks, there can subsequently be an orbit crossing (Figure \ref{fig:tad_pass}). Alternatively, a binary capture and merger (Figure \ref{fig:tadpole_to_merger}).

We can see why a tadpole breaks by plotting the evolution of the semi-major axis difference, $|a_1-a_2|$, and the distance $r$ between the masses, as shown in the bottom left panel of Figure \ref{fig:tad_pass}. Observe that in a tadpole, although the orbital distances $|a_1-a_2|$ is far less than the Hill radius, the instantaneous distance $r$ between the two BHs is always greater than the Hill radius. The tadpole is broken once the libration amplitude of the tadpole is large enough that close encounters with distances $\sim R_H$ occur. In the case where the encounter impact parameter is small enough, the result of energy losses from gas torques and gravitational wave emissions is great enough to form a binary (Figure \ref{fig:tadpole_to_merger}). On the other hand, tadpole breaking can result in the formation of a new tadpole (with a different initial libration period), or result in a orbit crossing. Our simulations indicate that mergers are the more typical outcome after the tadpole becomes unstable (see section \ref{sec:analytic_presc}), generally leading to a merger at a location well separated from the encounter location of $m_1$ and $m_2$.

\section{ANALYTIC PRESCRIPTIONS AND BRANCHING FRACTIONS}
\label{sec:analytic_presc}
In this section we consider the branching fractions of each of the outcomes described in Section \ref{sec:Results} occurring for a given encounter of BHs in an AGN disc. After numerically exploring the different outcomes via simulations, we apply an analytical model for resonance capture to infer the branching ratios of each outcome (and the probability for mergers) for different AGN discs. While we focus on GW190814, we perform a limited exploration of different mass ratios in section \ref{sec:mass_ratio_vary} \cite[also see][for complimentary analytical treatment]{epstein-mmr2025}.

\subsection{Mean motion resonance capture} \label{5.1}

\cite{batygin2015capture} proposed that resonance capture occurs for differentially migrating planets with period ratio $P_2/P_1=k/(k-1)$ if the resonance libration period of the $k:k-1$ resonance, $\tau_{\rm lib}$, is much less than the characteristic time, $\Delta t_{\rm res}$, for the system to cross the resonance width. This leads to the resonance capture condition: 

\begin{equation}
    \mathscr{B} \equiv \frac{\tau_{\rm{lib}}}{\Delta t_{\rm res}} \approx \frac{P_2}{\tilde{\tau}_{\rm mig}}  \left(  \frac{M_{\rm SMBH}}{m_1+m_2}  \right)^{4/3} \frac{1}{4 k^{2/9} (\sqrt{3} |f^{(1)}_{\rm{res}}|)^{4/3}} \ll 1,
    \label{eq:resonance_capture_con}
\end{equation}

\noindent
where $\tilde{\tau}_{\rm mig}^{-1}=\tau_{a,1}^{-1}-\tau_{a,2}^{-1}$ is the differential migration rate, and $f^{(1)}_{\rm{res}}$ follows the approximate scaling relation $f^{(1)}_{\rm{res}} \approx -0.46 - 0.802 k$ \citep{deck2013first}.

\begin{figure}
    \centering
    \includegraphics[width=0.48\textwidth]{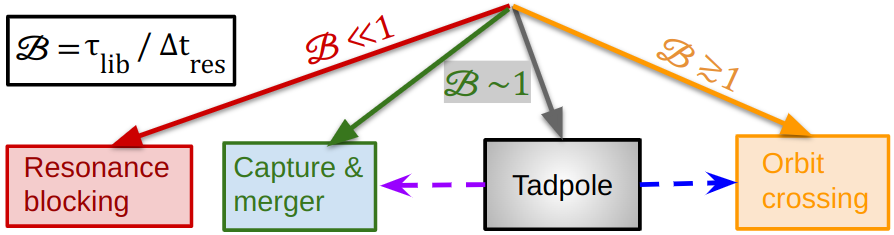}
    \caption{Flowchart of the possible outcomes. The key parameter is the ratio of the libration time $\tau_{\rm lib}$ to the resonance crossing time $\Delta t_{\rm res}$,  $\mathscr{B}=\tau_{\rm lib}/\Delta t_{\rm res}$, and is defined in Eqn. \ref{eq:resonance_capture_con}. For small $\mathscr{B}$, the libration is much faster than resonance crossing and capture into resonance is guaranteed. For $\mathscr{B}\gtrsim 1$, the libration is too slow, so orbit crossing is more likely. For $\mathscr{B} \sim 1$, the timescales are comparable and close encounters are more likely, which can lead to capture and merger or to a transient `tadpole' orbit, which leads to a binary capture or orbit crossing (dashed lines). The details and examples of each individual outcome are discussed in the rest of this section.}
    \label{fig:flowchart}
\end{figure}

This applies to any integer $k>1$, which would guarantee a resonance capture for a large enough $k$, so we impose an additional condition that bounds $k$. To ensure stability, the orbits of the sBHs need to be separated by a critical distance $R_c = \zeta \cdot  R_H$, where $R_H=a_2[(m_1+m_2)/3M_{\rm SMBH}]^{1/3}$ is the Hill radius. For $\Delta a = a_2 - a_1 = a_{2}\left[1-\left( (k-1)/k\right)^{2/3}\right]$ smaller than $R_c$ the system is unstable and may be captured into a binary. The value of $\zeta$ depends on details of the energy loss processes causing the capture. Detailed hydrodynamical simulations show that gas-assisted capture and mergers tend to be retrograde \citep{row23, li23agn}, which is also where the binary is most stable up to $\zeta=1$, while prograde orbits are unstable for $\zeta=0.5$ \citep{gri17}. We thus take $\zeta=1$ for a conservative estimate since the BH orbits around the SMBH will not be Keplerian once $\zeta<1$ and the resonance will be broken. We will later see that the exact choice of $\zeta$ is not important for our results. The maximum value $k$ for stable orbit is estimated as:

\begin{align}\label{eq:kmax}
    k_{{\rm max}}&\approx\frac{2}{3\zeta}\left(\frac{3M_{\rm SMBH}}{m_1+m_2}\right)^{1/3}=150 \zeta^{-1} \left(\frac{M_{{\rm SMBH}}}{10^{8} M_{\odot}}\right)^{1/3},
\end{align}
where we used a Taylor expansion for $(1+(R_c/a_2))^{3/2}\approx 1 + 3 R_c/(2a_2)$.

It is also easy to see that if the opposite condition is true, $\mathscr{B} \gg 1$, and hence resonance capture has a vanishingly small probability of occurring, then an orbit crossing without a binary capture occurs with high probability. To see this, consider that during an orbit crossing (where $a_1-a_2$ switches signs), a binary capture will occur if a conjunction of the two BH leads to a small enough instantaneous distance that energy loss from gas interactions and gravitational wave emission leads to a binary forming. Thus, if the time between conjunctions, $\Delta t_{\rm conj} = 2 \pi a_2/ \Delta v$ (where $\Delta v = 1/2 \sqrt{GM/a_2^3} R_{\rm H}$ is the approximate relative velocity of Keplerian orbits displaced by $\Delta a=R_{\rm H}$), is smaller than the Hill crossing time $\Delta t_{\rm H}=2 R_{\rm H} / \Delta v_{\rm mig}=2 R_{\rm H} \tilde{\tau}_{\rm mig} / a_2$ for the outer object to migrate through the Hill sphere of the inner object, then an orbit crossing will occur with high probability. Thus, an orbit crossing will occur (without binary capture) with high probability if the ratio $\Delta t _{\rm H} / \Delta t_{\rm conj} \ll 1$, which is equivalent to the condition:

\begin{equation}\label{eq:hill_pass_by}
    \frac{P_{\rm 2}}{ \tilde{\tau}_{\rm mig}} \gg \left(\frac{ R_{\rm H}}{ a_2 }\right)^2= \left(\frac{m_1+m_2}{3 M_{\rm SMBH}} \right)^{2/3}.
\end{equation}

\begin{figure}
    \centering
    \includegraphics[width=0.5\textwidth]{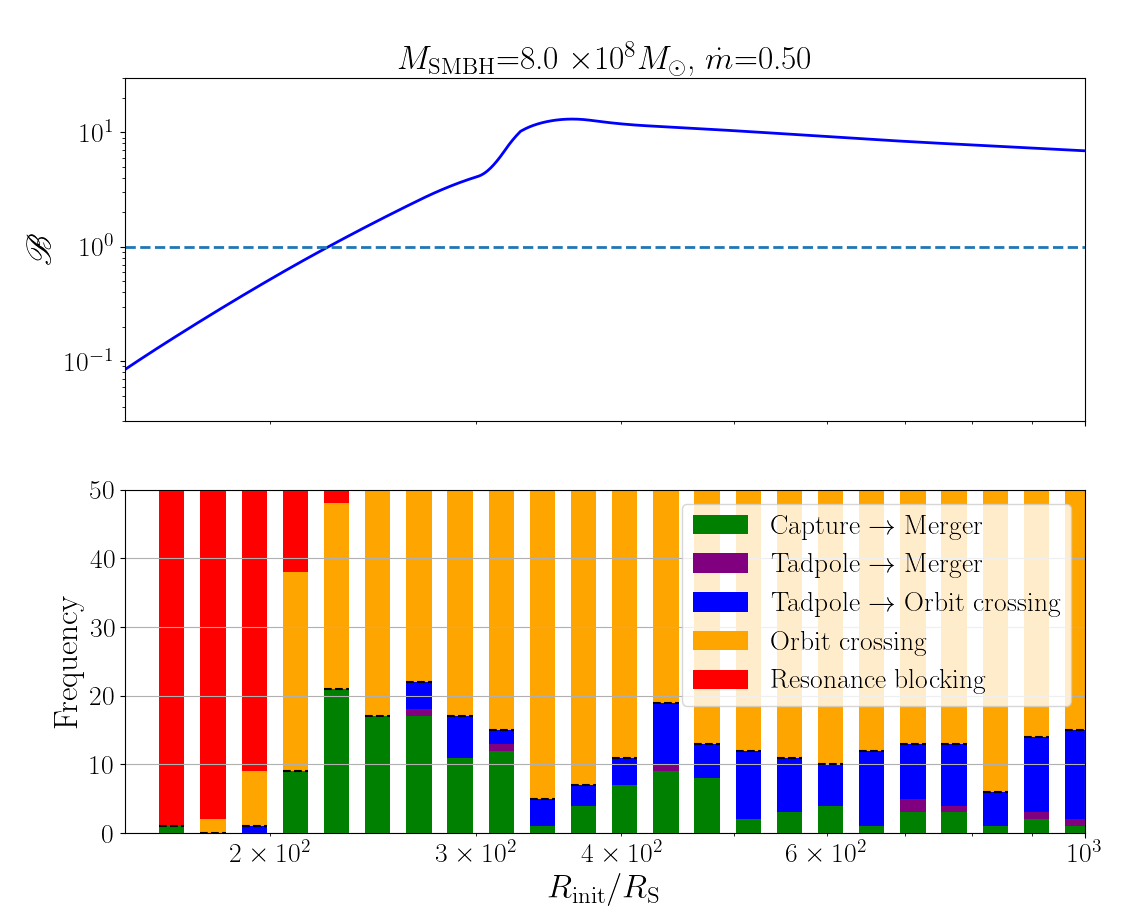}
    \caption{Branching fractions of encounter possibilities of $m_1$ and $m_2$ differentially migrating in the disc $(M_{\rm SMBH}/M_{\odot}, \dot{m}, \alpha)=(8 \times 10^8, 0.5, 0.01)$ and encountering one another at various radii from the central SMBH. Note that the initial radii where encounters tend to lead to mergers with high probability correspond to locations where $\mathscr{B} \sim 1$ (plotted above), while regions with $\mathscr{B} \ll 1$ tend to form resonances preventing merger, and regions where $\mathscr{B} \gg 1$ favor orbit crossings.}
    \label{fig:bat_8e8}
\end{figure}
Compareing this to $\mathscr{B} \gg 1$, with $\mathscr{B}$ given by Eqn. \ref{eq:resonance_capture_con} with $k=k_{\rm max}$ (Eqn. \ref{eq:kmax}):

\begin{equation}
    \frac{\tau_{\rm lib}}{\Delta t_{\rm res}} \approx \frac{P_2}{\tilde{\tau}_{\rm mig}} \left(\frac{m_1+m_2}{3 M_{\rm SMBH}}\right)^{-2/3}  \left[0.07 \left(\frac{m_1+m_2}{3 M_{\rm SMBH}}\right)^{-4/27} \zeta^{14/9} \right] \gg 1,
\end{equation}
which yields the same scaling as Eqn. \ref{eq:hill_pass_by}, up to the factor $\left[0.07 ((m_1+m_2)/3M_{\rm SMBH})^{-4/27} \zeta^{14/9} \right]$ that varies very little over our parameter space. This suggests that the outcome of converging orbits of black holes migrating in an AGN disc can be predicted throughout the AGN disc by computing $\mathscr{B}$, which is determined only by the BH masses and the local disc properties.

In particular, when $\mathscr{B} \ll 1$, slow convergent migration results in a resonance that prevents close encounters, while $\mathscr{B} \gg 1$ ensures a high probability of orbit crossing due to rapid differential migration. In the range $\mathscr{B} \sim 1$, we expect a high rate of close encounters, and hence the formation of either binary captures or co-orbital resonances, with the probability of such an outcome peaking around $\mathscr{B} = 1$.

Figure \ref{fig:flowchart} shows the flowchart of the possible outcomes and their interconnections. Each outcome occurs probabilistically depending on the AGN disc properties at the encounter location.  While we provide a convenient fitting formulae for the total capture into a binary or a tadpole orbit in the next section (solid lines), the estimate of the branching ratio outcomes of the tadpoles after destabilization (dashed blue and purple lines) are not modeled. More accurate estimates requires additional simulations and will be addressed in future work. Despite this, we will see in sections \ref{sec:merger_branch_disc} and \ref{sec:mass_ratio_vary} that without modeling this branching ratio, our fitting formula still gives good estimates of the binary capture fractions at trap locations.

\subsection{Capture probability prescription}\label{sec:merger_prob_prescription}

\begin{figure*}
    \centering
    \includegraphics[width=1.0\textwidth]{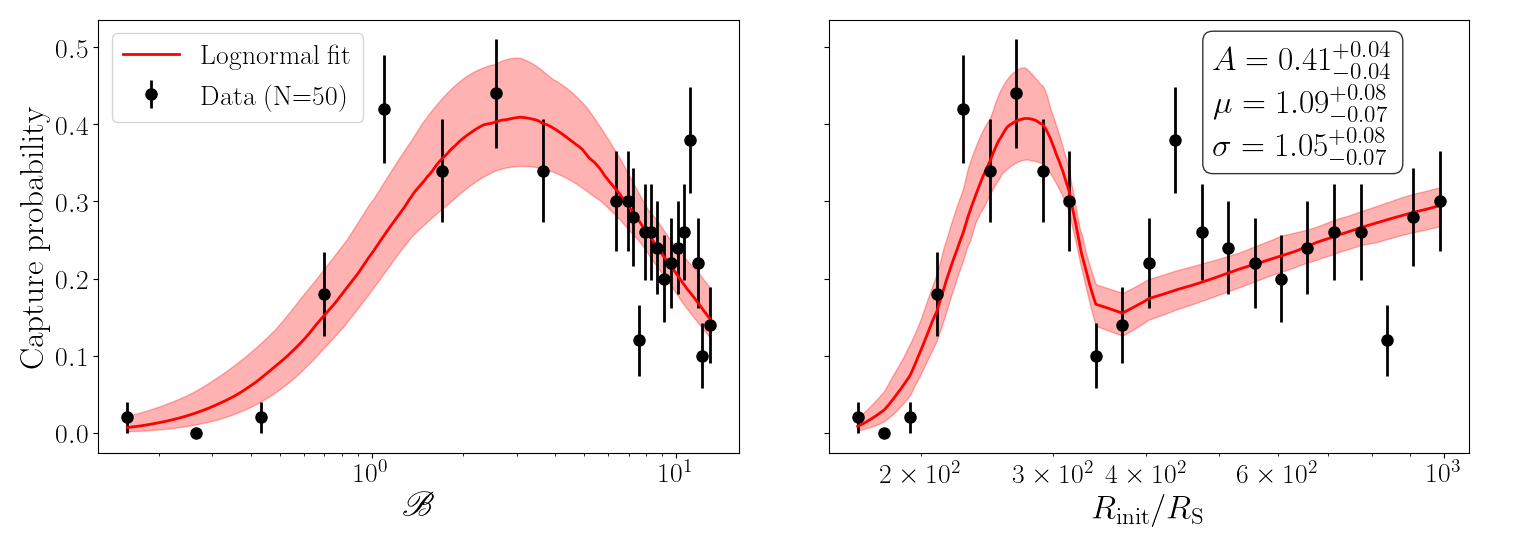}
    \caption{Empirical fit of capture fraction, displayed on the left as a function of $\mathscr{B}$. Data points are obtained by computing the fraction of simulated events at each $R_{\rm init}$ (out of $N=50$) that result in either a binary or tadpole capture. The red curve shows the best fit of our empirical model (Eqn. \ref{eq:merger_prob}). The data points show $1\sigma$ error bars, assuming the error due to a finite sample follows a Binomial distribution. The shaded region shows the 95 per cent confidence intervals on the lognormal fit. The plot on the right is the same as the left, except the capture probability is plotted against $R_{\rm init}/R_s$.}
\label{fig:merger_fit}
\end{figure*}

We next study the branching fractions of each orbital encounter outcome as a function of the distance from the SMBH. To do this, we initialise the masses $m_1=23.2 M_{\odot}$ and $m_2=2.59M_{\odot}$ in the disc $(M_{\rm SMBH}/M_{\odot},\dot{m},\alpha)=(8 \times 10^8, 0.5, 0.01)$ (which contains no traps) at starting positions $r_1$ and $r_2$, where $r_1=1.05 r_2$ and $r_2$ consists of $23$ points logarithmically spaced between $10^2 R_S$ and $10^3 R_S$, for a total of $1150$ simulations. Outside this radial range, the differential migration rate of the BHs becomes comparable to the expected lifetime of a AGN disc. We run each simulation $N=50$ times, with the initial orbital variables sampled randomly from the distributions described in section \ref{sec:coupling_codes}. We tabulate the number of encounters that result in binary capture (green), tadpole formation (blue or purple depending on the end state), Hill-limit resonance blocking (red), or an orbit crossing without further interaction (orange).

Simulations are stopped and classified as "Capture $\rightarrow$ Merger" if the spatial distance between the bodies dips below $d < 0.01 \rm{AU}$. We do not evolve the binaries further than this in our suite of simulations, because tighter binaries quickly become highly relativistic, and it is computationally expensive to evolve further. However, binaries that have hardened to this small of a radius should be in the GW dominated regime, and therefore merge rapidly (as shown in Figure \ref{fig:orb_evo_merger}). Simulations are classified as "Resonance blocking" if the more rapidly migrating $m_1$ never crosses the orbit of $m_2$, and the period ratio remains bound between two first order MMRs for the final third of the simulation run, i.e., for some $k \in \mathbb{N}$:

\begin{equation}
    \frac{k}{k-1} < \frac{P_1}{P_2} < \frac{k+2}{k+1}.
\end{equation}

Simulations are classified as tadpoles if the period ratio of the inner and outer binary goes through at least five periods of oscillation about $P_2/P_1=1$, without ever having a close enough encounter for the binding energy of the $m_1$--$m_2$ system to become negative (a necessary, but not sufficient, condition for binary capture). The tadpoles are then classified as "Tadpole $\rightarrow$ Merger" if after the transient tadpole state leads to a binary capture upon destabilization, otherwise the tadpole is classified as "Tadpole $\rightarrow$ Orbit Crossing". If the orbit of $m_2$ crosses that of $m_1$, and no other condition is met, then the simulation is classified as an "orbit crossing".

The top panel of Figure \ref{fig:bat_8e8} shows $\mathscr{B}$ as a function of $R_{\rm S}$ for our disc model. The results of the simulation are shown in the bottom panel of Figure \ref{fig:bat_8e8} as a bar chart. We observe that the locations where $\mathscr{B} \ll 1$ (in the inner disc) lead mostly to resonance blocking, while locations where $\mathscr{B} \gtrsim 1$ result in orbit crossings. Regions where $\mathscr{B} \sim 1$ lead to a high number of mergers or tadpoles orbits forming, with the remainder resulting in an orbit crossing. This qualitatively fits the MMR resonance capture criterion covered in the previous section.

\begin{figure*}
    \centering
    \begin{subfigure}[b]{0.55\textwidth}
        \includegraphics[width=1\textwidth]{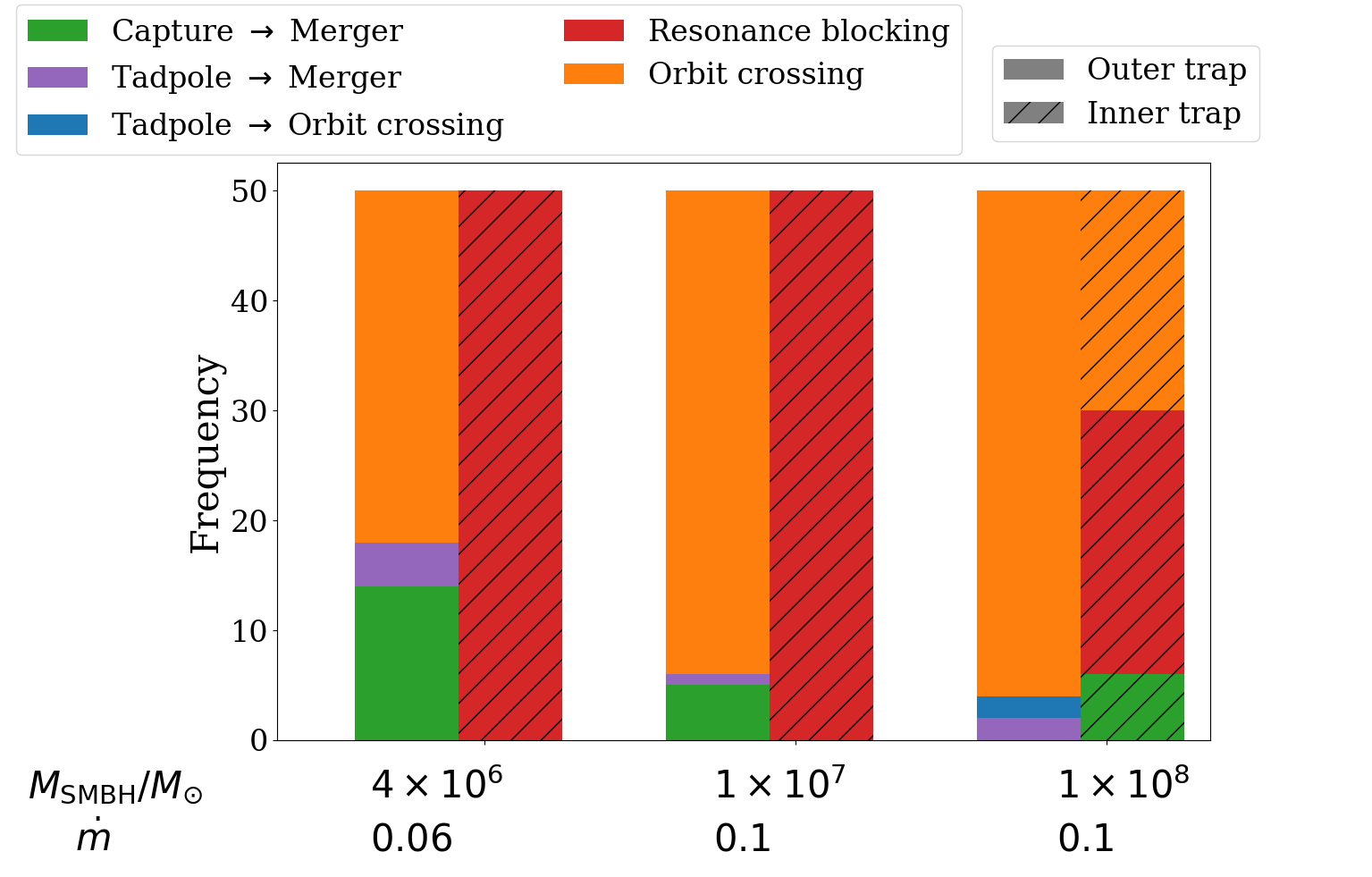}
    \end{subfigure}
    \hspace{0.05cm}
    \begin{subfigure}[b]{0.42\textwidth}
        \raisebox{-0.1cm}{\includegraphics[width=\textwidth]{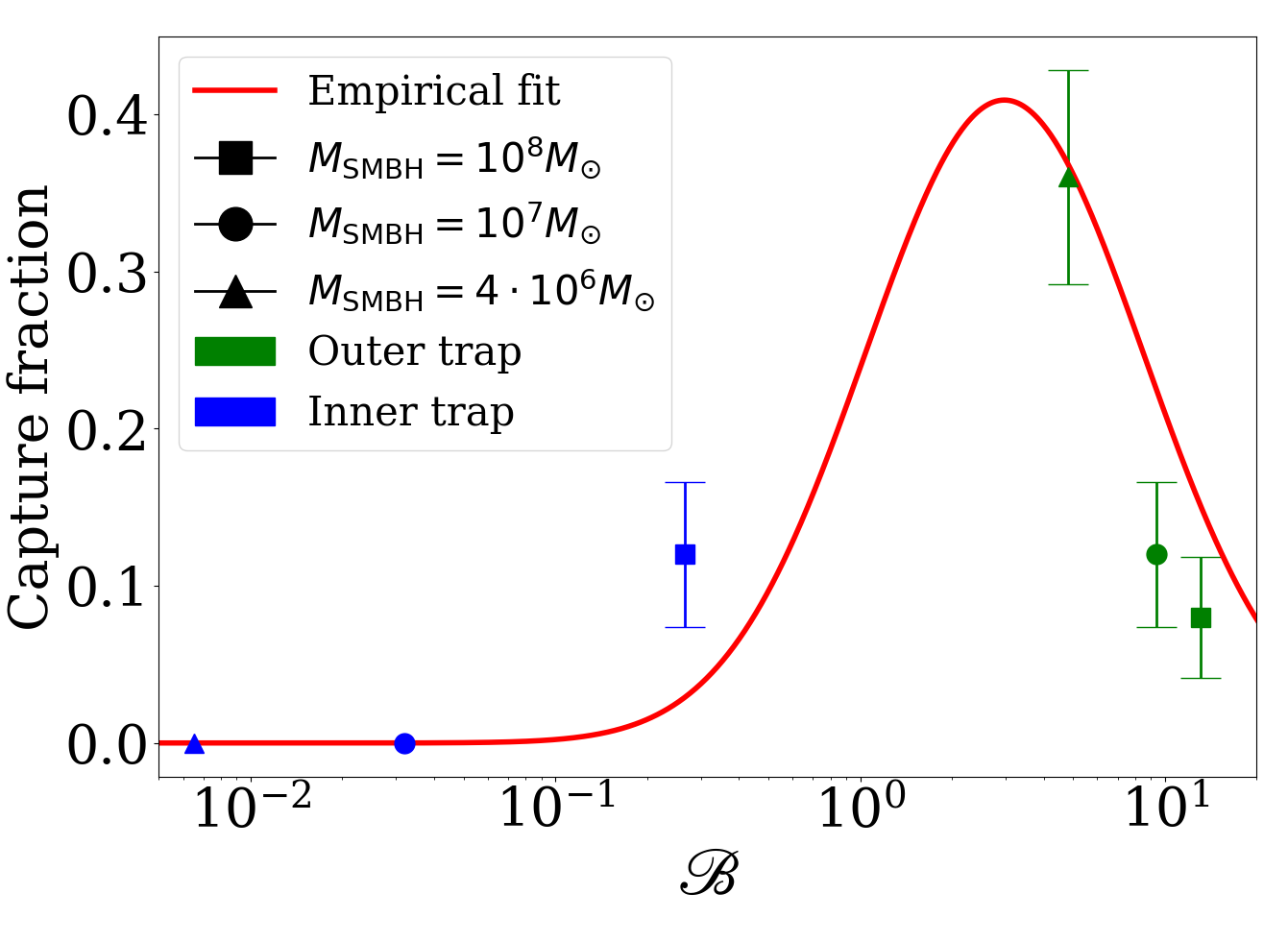}}
    \end{subfigure}
    \caption{Results of 50 simulations of encounters at each trap of each of our three discs of interest. The bar plot on the left shows the branching fractions for our three discs containing traps, with the the hatched histogram showing the outcome at the inner trap, and the unhatched histogram show the outer trap. The Figure on the right shows fraction of captures (binary or tadpole capture) resulting from the 50 encounters at each trap, compared with the fraction estimated when we compute $P(\rm{capture}|\mathscr{B})$ (Eqn. \ref{eq:merger_prob}) using our best fit values from section \ref{sec:analytic_presc}.}
    \label{fig:outcome_fractions_discs}
\end{figure*}

Motivated by the required asymptotic behavior of the capture probability as a function of $\mathscr{B}$, we can use these simulations to construct an empirical fitting formula for the capture probability as a function of $\mathscr{B}$. Here, and henceforth, we will refer to the case of either binary capture, or capture into a tadpole orbit, as a `capture'. We find that a normal distribution in the parameter $\ln{\mathscr{B}}$ has the correct asymptotic values in the PDF, so we fit the capture fraction as a function of SMBH separation to a Gaussian PDF:

\begin{equation}\label{eq:merger_prob}
    P(\rm{capture}|\mathscr{B})=A \exp{\left(-\frac{(\ln\mathscr{B}-\mu)^2}{2 \sigma^2}\right)}
\end{equation}
where the parameters $A, \mu, \sigma$ are parameters to be fitted from simulations. The parameter $\mu$ has the interpretation of the value of $\ln{\mathscr{B}}$ where the capture probability is maximum, i.e. when  $\mathscr{B}=e^{\mu}$, and $A$ is the maximum capture probability for this condition. Finally, $\sigma$ determines how fast the capture probability decays in the limit of $\ln{\mathscr{B}} \rightarrow \pm \infty$ ($\mathscr{B} \ll 1$ and $\mathscr{B} \gg 1$). The best fit for our data 
is shown in Figure \ref{fig:merger_fit}, with best fit parameters $A=0.41,\mu=1.10$, and $\sigma=1.05$, corresponding to $\mathscr{B}\approx3$\footnote{Note that the value of $\mu$ in our fit is dependent on our choice of $\zeta$, changing out choice of $\zeta$ in Eqn. \ref{eq:kmax} alters the best fit value of $\mu$. However, as long as there is consistency between the value of $\zeta$ chosen during fitting and when computing $\mathscr{B}$, the calculated $P(\rm{capture}|\mathscr{B})$ will be independent of $\zeta$.}.  

We will use the best-fit parameters in the next sections, verifying that the empirical formula generalizes, giving reasonable results for different discs and with different mass ratio encounters.

\subsection{Merger rates across AGN disc parameter space}\label{sec:merger_branch_disc}

We now turn to studying encounters at the trap locations of our three discs containing traps. For each disc, we run 50 simulations of $m_2$ migrating into the trap location of $m_1$, enumerating the possible outcomes of encounters as in the previous section. Each disc contains two traps (Figure \ref{fig:type_2_torques_plots}), an inner trap with radius $a \sim 10^2R_{\rm S}$ from the central SMBH, and an outer trap with $a\sim10^3R_{\rm S}-10^4 R_{\rm S}$. The left panel of Figure \ref{fig:outcome_fractions_discs} shows the outcomes for encounters at each trap. We observe that at the outer traps orbit crossing is the most common outcome, while at inner traps resonance formation dominates. We also see that the lowest luminosity disc, $(M_{\rm SMBH}/M_{\odot}, \dot{m}, \alpha)=(4 \times 10^6, 0.06, 0.01)$, has by far the highest capture probability at the outermost trap, because  the relative migration rate between $m_1$ and $m_2$ is the lowest.

\begin{figure*}
    \centering
    \begin{subfigure}[b]{0.49\textwidth}
        \centering
        \includegraphics[width=1\textwidth]{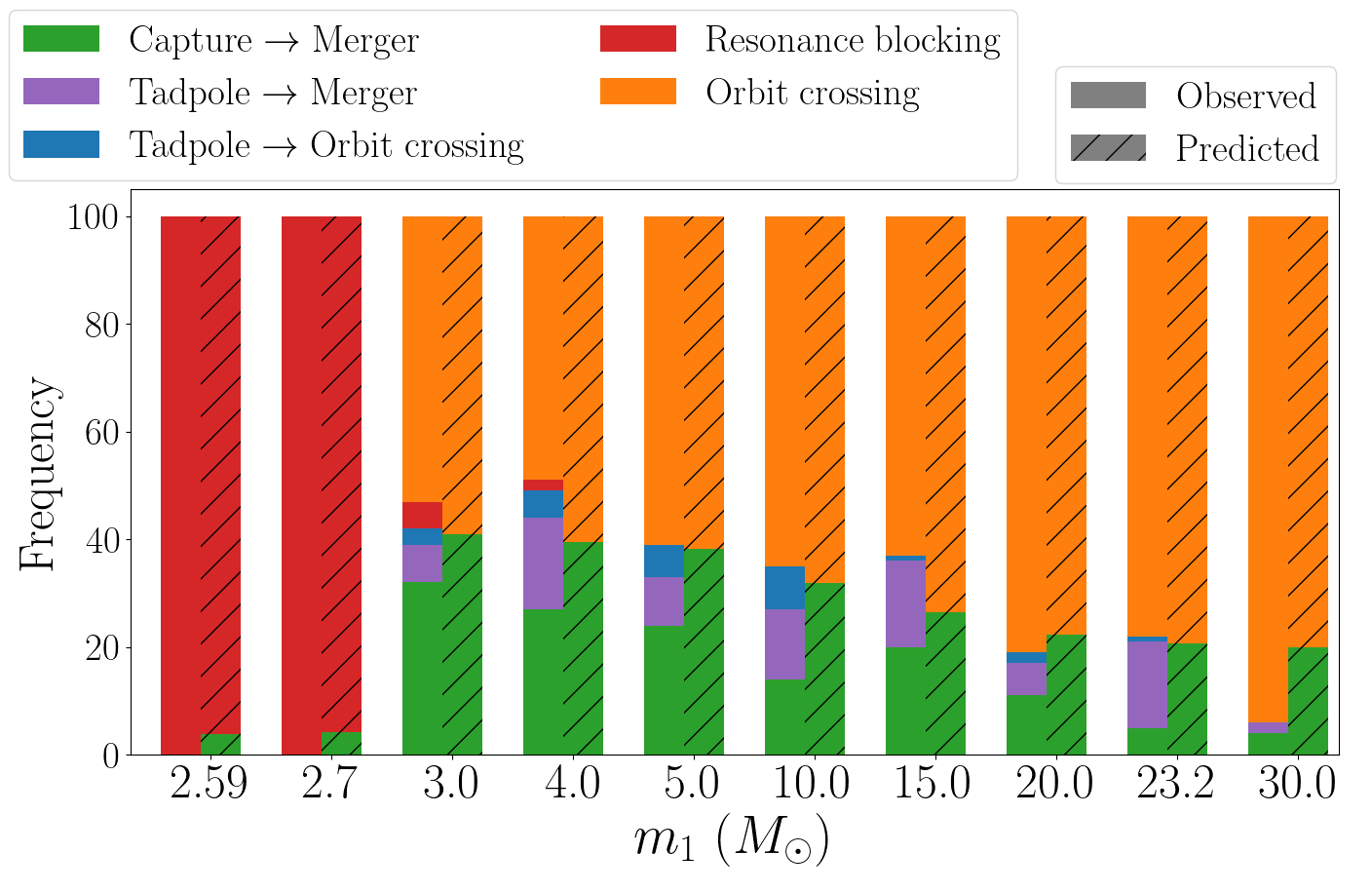}
    \end{subfigure}
    \hspace{0.1cm}
    \begin{subfigure}[b]{0.49\textwidth}
        \centering
        \includegraphics[width=1\textwidth]{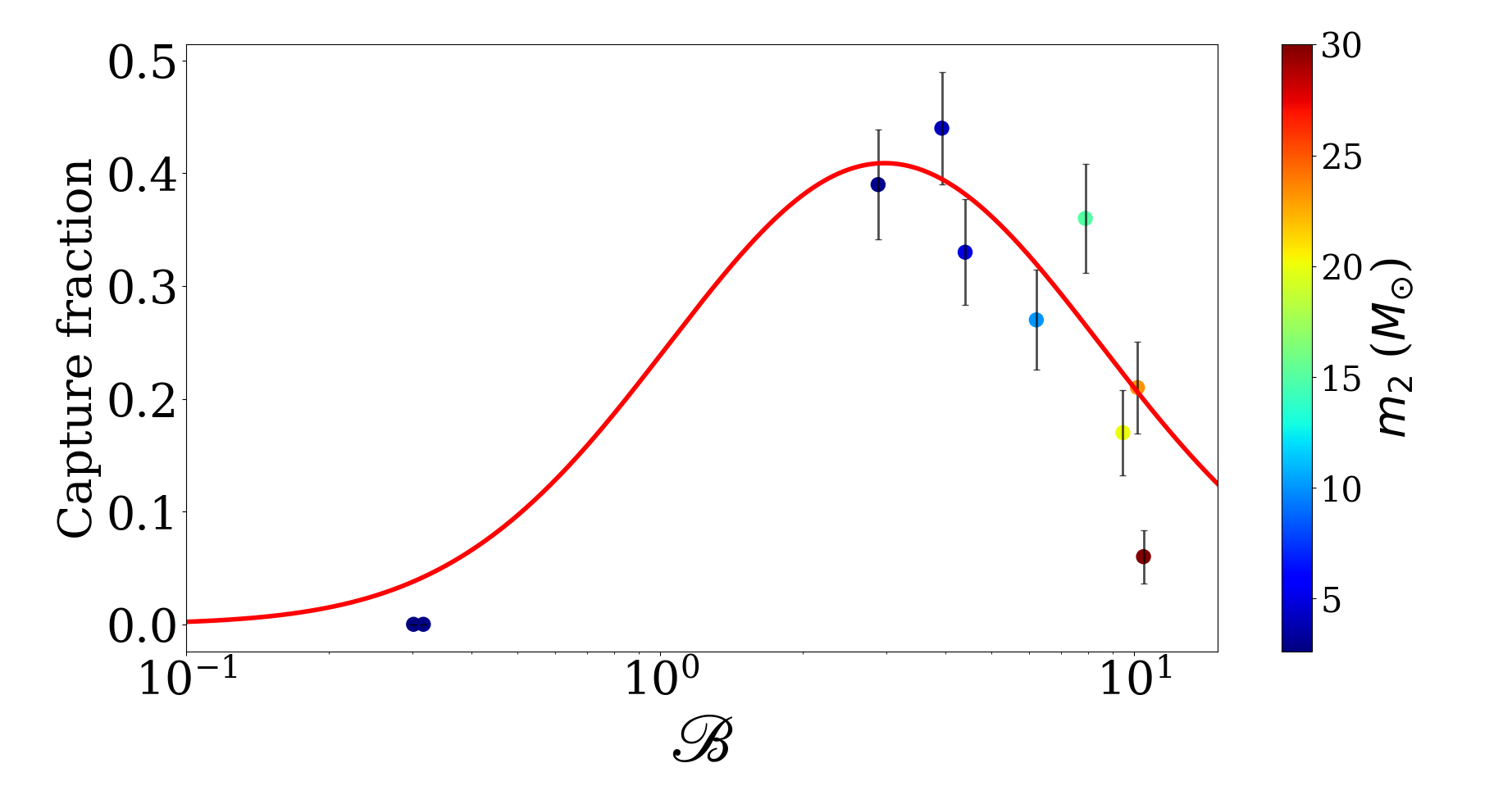}
    \end{subfigure}
    \caption{Results of 100 simulations of encounters at the trap location of $m_2=2.59 M_{\odot}$ in the $(M_{\rm SMBH}/M_{\odot}, \dot{m}, \alpha)=(10^7, 0.1, 0.01)$ AGN disc, for various different value of $m_1$. The bar chart on the left shows the branching fractions for the simulations at each simulated value of $m_1$.The unhatched histograms enumerate the outcomes of each of the 100 simulations, as in Figure \ref{fig:outcome_fractions_discs}, while the adjacent hatched plots for each $m_2$ show the fraction of expected captures. This expected capture fraction is computed from the local AGN disc values of $\mathscr{B}$ (which is also a function of $m_2$) and our fit from section \ref{sec:analytic_presc}. The right figure shows the capture fraction of each simulated $m_1$ as a function of $\mathscr{B}$, comparing the results of the simulation to the best fit curve for $P(\rm{capture}|\mathscr{B})$.}
    \label{fig:branching_frac_by_msecondary}
\end{figure*}

The right panel of Figure \ref{fig:outcome_fractions_discs} shows the capture fractions as a function of $\mathscr{B}$ at each trap. We can see how the capture fractions observed in our simulations depend on the value of $\mathscr{B}$ for an $m_1$-$m_2$ encounter. We observe that the inner traps have a low value of $\mathscr{B}$, explaining the high number of resonances, while the outer traps have a higher value of $\mathscr{B}$, generally leading to a higher number of captures and orbit crossings. We also see that the empirical model, fitted to our $(M_{\rm SMBH}/M_{\odot}, \dot{m}, \alpha)=(8 \times 10^8, 0.5, 0.01)$ disc simulations, provides good estimates for the capture probability at the trap locations for different discs. 

\subsection{Dependence of trap merger rate on mass ratio}\label{sec:mass_ratio_vary}

In order to test the robustness of our prescription, we fix the AGN disc to  $(M_{\rm SMBH}/M_{\odot}, \dot{m}, \alpha)=(10^7, 0.1, 0.01)$ and $m_1$, but vary $m_2$. We run 10 different values of different $m_2$ at $10^4 R_s$ between $2.59 M_{\odot}$ and $30 M_{\odot}$, each one for $100$ times at the trap location, $a=9.4 \times 10^3 R_s$.

The results are shown in Figure  \ref{fig:branching_frac_by_msecondary}, where we tabulate the outcomes of each encounter in a histogram (left). Overall, we find good agreement between the prediction using the empirical fit and the simulations. The model correctly predicts resonance blocking preventing mergers when $m_1 \approx m_2$, with a peak in capture probability around $m_1/m_2=1/2$, followed by a decay as the mass ratio decreases further. Note that the value of $\mathscr{B}$ experiences rapid jump from $\mathscr{B}\approx 0.3$ to $\mathscr{B}\approx 3$ as $m_2$ goes from $2.7 M_{\odot}$ to $3 M_{\odot}$, leading to a sharp increase in capture probability (right panel of Figure \ref{fig:branching_frac_by_msecondary}). This results from partial gap opening leading to the outermost trap disappearing when $m_1\approx2.8 M_{\odot}$, drastically increasing the migration rate at the trap location.

\begin{figure*}
    \centering
    \begin{subfigure}[b]{\textwidth}
        \includegraphics[width=0.95\textwidth]{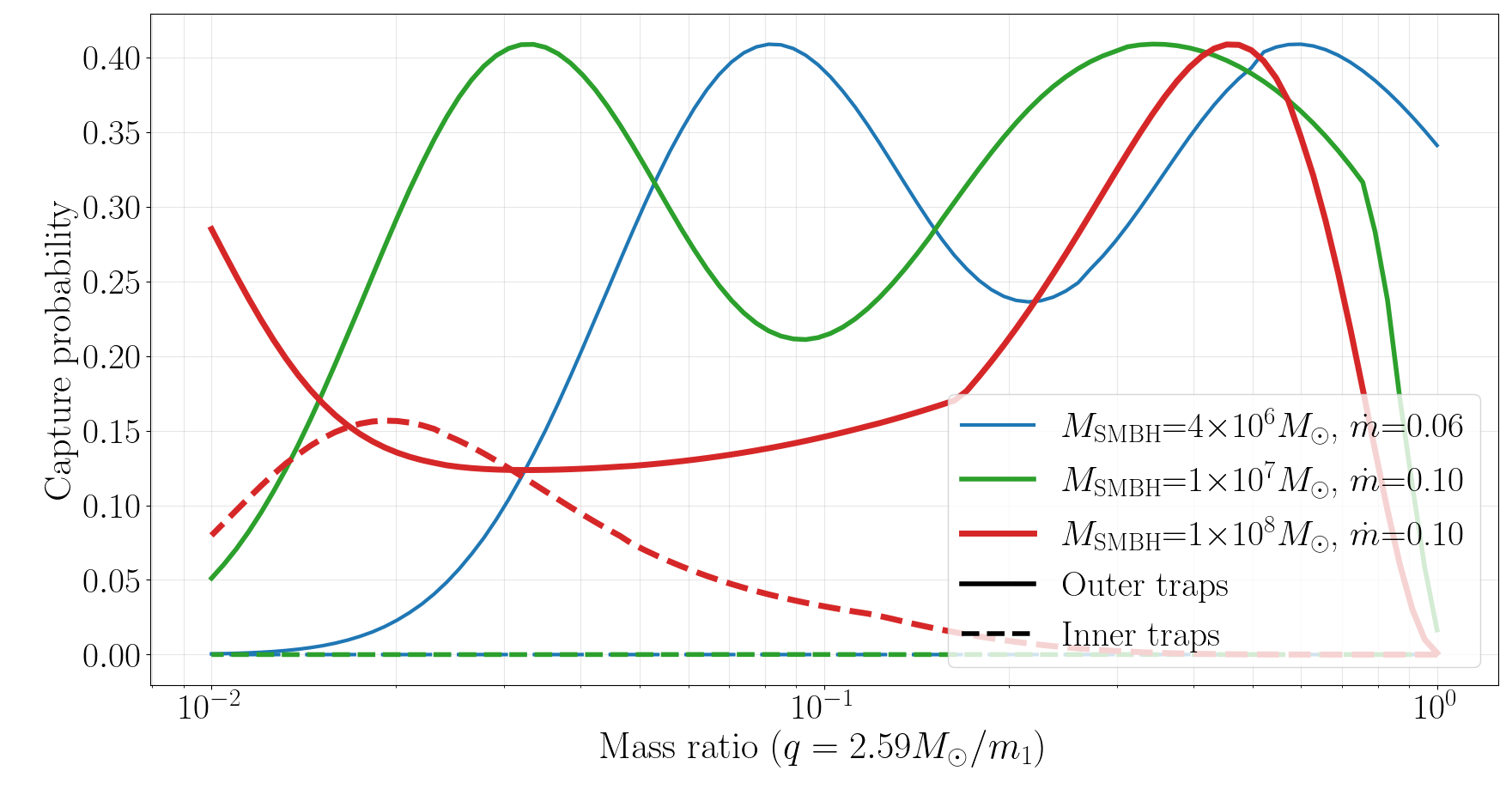}
    \end{subfigure}
    \hfill
    \begin{subfigure}[b]{\textwidth}
        \includegraphics[width=\textwidth]{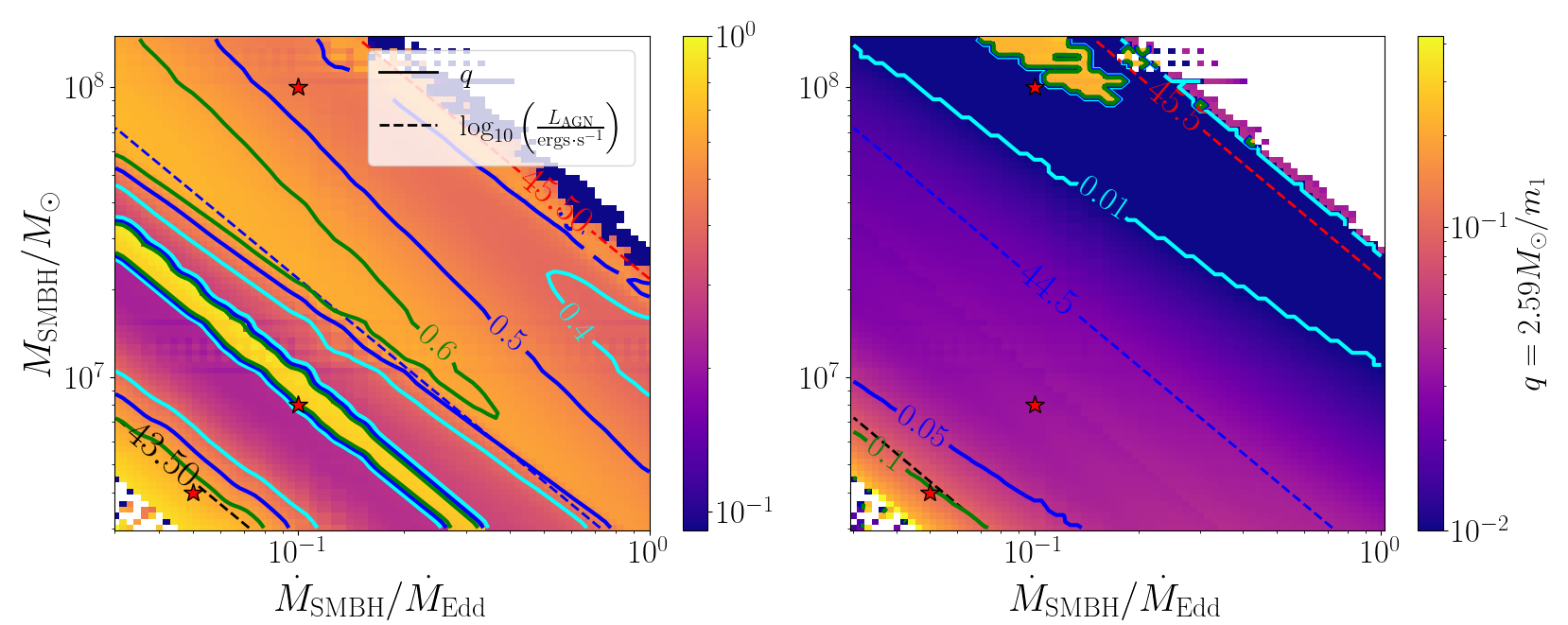}
    \end{subfigure}
    \caption{Top: Capture probability as a function of mass ratio at trap locations of $m_2=2.59 M_{\odot}$, for each of the discs with traps. Obtained by applying our empirical model to the three discs inner and outer trap locations, for various value of $m_1$ in the interval $[m_2, 100m_2]$. The solid (dashed) lines show the capture probabilities at the outer (inner) trap locations.
    Bottom: Heat map showing the mass ratio $q=2.59 M_{\odot}/m_1$ for $m_1$ that maximizes the capture probability at the outer trap of $m_2=2.59 M_{\odot}$ for all discs of our AGN parameter space. Since there are two peaks (as seen in the top figure), the left hand plot shows the higher $q$ peak, while the right hand plot shows the lower $q$ peak (where $m_1$ is in the deep-gap limit). Contour lines for constant $q$ are shown as solid lines, and contours for fixed luminosity are shown as dashed lines.}
    \label{fig:q_param_space}

\end{figure*}

Using our empirical mode, we can see how our representative AGN discs have capture probabilities that peak at different mass ratios $q=m_2/m_1$ for a fixed $m_2=2.59 M_{\odot}$. The top panel of Figure \ref{fig:q_param_space} shows $P(\rm{capture})$ at the trap locations of our three discs, evaluated as a function of $m_1$ (with $m_2=2.59 M_{\odot}$ fixed). We observe clear peaks in the capture probability that vary with the discs, but generally have a similar doubly peaked structure. The first peak occurs when the differential migration torques can overcome the resonance, dipping back down once the differential torques are too large and orbit crossing dominates again. 

The second peak, on the other hand, is a consequence of gap opening for large BHs. When $m_1\sim m_{\rm gap}$, gap opening leads to a greatly reduced local gas density, significantly reducing the migration torques. In the limit that $m \gg m_{\rm gap}$, the factor $\exp(-K/20) \rightarrow 0$, and hence (Eqs. \ref{eq:torque_tot} and \ref{eq:type_II_torque}) the total torque goes to:

\begin{equation}
    \Gamma_{\rm tot}= \frac{C_{\rm L}}{1+K/25} \Gamma_0,
\end{equation}
i.e the only torque contribution is from the Lindblad torque. Since $K=25 (m/m_{\rm gap})^2 \propto m^2$, and $\Gamma_0 \propto m^2$, the total torque $\Gamma_{\rm tot}$ is independent of mass, as expected in classical type II migration. Thus in the limit of large masses we return to the \cite{bellovary2016migration} picture, though the migration timescales will be extremely large, with $\tau_{\rm mig} \propto m$.

For a large enough sBH mass, the differential migration rate of the larger BH can therefore decrease back down to the value of the smaller mass, increasing the merger probability. Eventually, when the mass is very large, resonance blocking becomes dominant again, leading to a merger probability decaying back to zero. However, the migration timescales of these very large mass black holes are much larger than the lifetime of an AGN, leading to a low encounter probability for these deep-gap objects. Though it is possible that these massive objects could result from multi-generation sBH mergers, if hierarchical mergers are efficient near a trap location. 

We further explore the critical values of $q_{\rm peak}=2.59M_{\odot}/m_1$ that maximize the capture probability by plotting a heat map showing the critical values of $q_{\rm peak}$ for a grid of AGN disc parameters $(M_{\rm SMBH}, \dot{m})$ in the bottom two plots of Figure \ref{fig:q_param_space}. Since there are generally two peaks in $q$, we plot the first (second) peak value of $q$ in the bottom left (right). In close analogy with the gap opening mass displayed in Figure \ref{fig:Mgap_hr}, we observe that the optimal $q$ values for mergers at a trap location are well parameterized by the AGN luminosity. For example, the AGN disc $(4\times 10^6 M_{\odot}, 0.06)$, our disc of interest in Figure \ref{fig:outcome_fractions_discs} that had the highest capture fraction at the outermost trap, lies close to the $q_{\rm peak}=0.1$ contour line in the bottom right panel (corresponding to a GW190814 like binary). This contour line, which closely aligns with the $L_{\rm AGN}=10^{43.5}\ \rm{erg}\ \rm{s}^{-1}$ contour, is therefore a region in the parameter space of AGN that could be conducive to producing GW190814 like systems.

\begin{figure}
    \centering
    \includegraphics[width=0.5\textwidth]{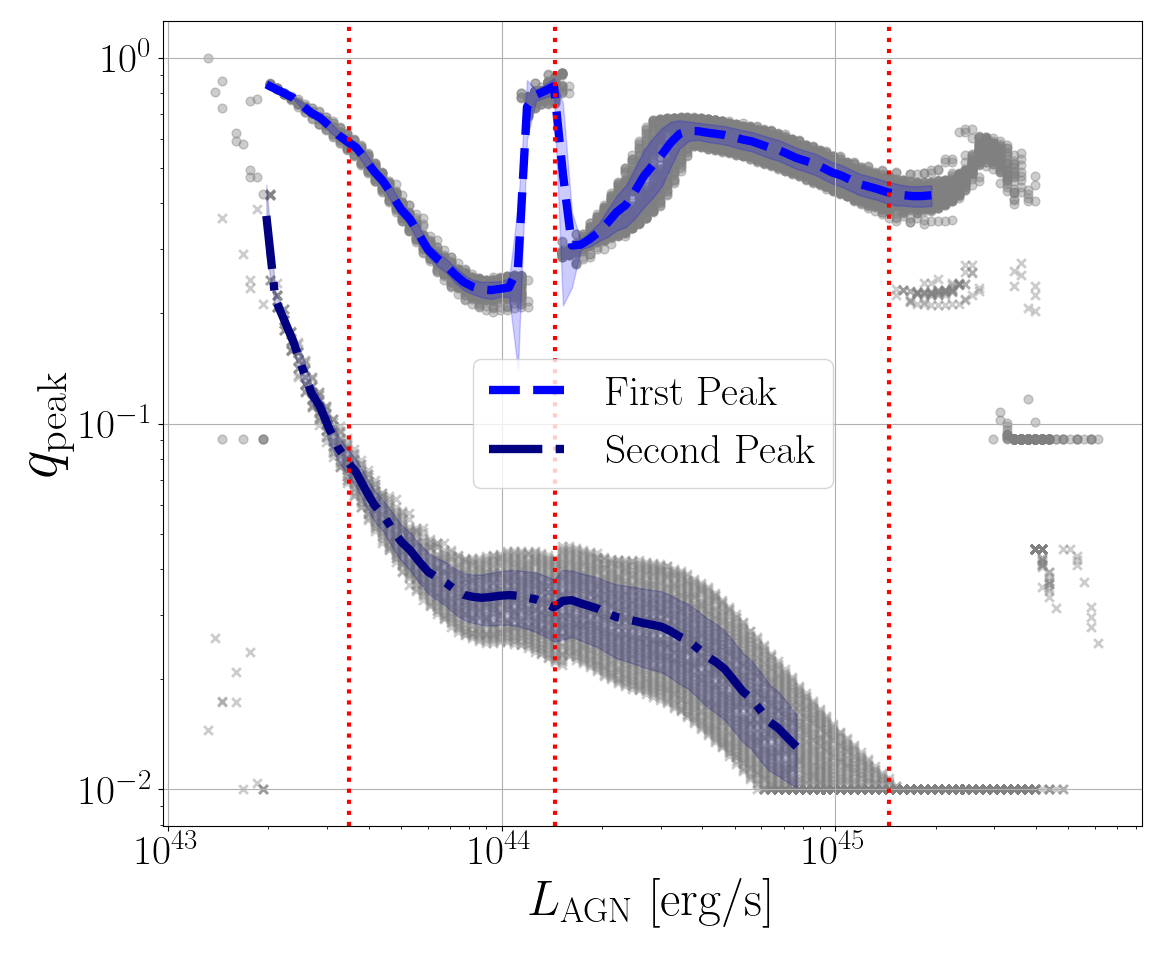}
    \caption{The same data as Figure \ref{fig:q_param_space}, but with the optimal $q$ value for each AGN disc $(M_{\rm SMBH}, \dot{M}_{\rm SMBH})$ parameterized by the AGN luminosity $L_{\rm AGN}$. The grey dots correspond to the first peak in the mass ratio $q_1$ for each disc, while the crosses correspond to the second peak $q_2$. The shaded regions show $1\sigma$ error bars about the mean value of $q_{\rm peak}$ for a given luminosity. The vertical red dashed lines correspond to the luminosities of each of our discs of interest (except the $M_{\rm SMBH}=8 \times 10^8 M_{\odot}$ disc, which does not have any traps). Values of $q_1<0.1$ and $q_2<0.01$ are not shown.}
    \label{fig:q-L_plot}
\end{figure}

Figure \ref{fig:q-L_plot} shows a scatter plot with all of the $q_{\rm peak}$ values for each disc, against the AGN discs luminosity $L_{\rm AGN}$ (Eqn. \ref{eq:L_AGN}). We see a strong correlation between the AGN luminosity and the two $q_{\rm peak}$ values. We see that the first peaks tend to produce mass ratios in the range $q\in(0.2,0.7)$ while the second peak has lower mass ratios. GW190814-like events are more rare, but can be produced due to the second peak for low luminosity AGNs around $L_{\rm AGN} = 10^{43.5}\ \rm erg\ s^{-1}$.

\section{DISCUSSION}
\label{sec:discussion}

In this paper we have performed N-body simulations of sBHs migrating in various AGN accretion discs, and identified the possible outcomes when pairs of sBH with masses consistent the GW190814 event encounter each other in the disc. We find several possible outcomes for these encounters: binary formation and subsequent merger, orbit crossing with little interaction, or the formation of a MMR - including a co-orbital (i.e. tadpole or horse-shoe) resonances. The branching fraction of each outcome during encounters depend on the AGN disc properties, sBH masses, and the location of the encounter in the disc. We demonstrate how this dependence can be explained through analytic models of resonance capture, enabling us to make predictions of the BH mass ratio that can merge both inside and outside of migration traps in AGN discs.

The analytical resonance capture condition (i.e. the ratio of the libration timescale to the migration pass-by time; Eqn. \ref{eq:resonance_capture_con}) shows that while BH systems with wide mass ratio range can result in mergers, the close encounter probability strongly depend on the radial distance from the central SMBH, and often \textit{not} highest at the migration trap location (which may be naively expected). This suggests that each disc contains a range of orbital radii where encounters between $m_1$ and $m_2$ are in the sweet spot of neither migrating slow enough to form MMR, nor fast enough for orbit crossing. Thus, the formation of MMR and mass dependent trap locations due to thermal torques and gap opening, will \textit{suppress} the rate of BBH mergers in AGN discs. This may mean that the merger rate of BHs from the AGN channel is lower than previous estimates suggest \citep{mckernan2020black,grobner2020binary,tagawa2020formation}.

\subsection{Astrophysical implications}

\subsubsection{GW190814}

In our simulations, we identify several cases where GW190814-like events can be produced in an AGN disc (mostly AGNs on the lower end of AGN luminosity of $L_{\rm AGN} = 10^{43.5}\ \rm erg\ s^{-1}$), so we cannot rule out the possibility of this event originating from the AGN channel. However, we have seen that an orbital encounter of highly unequal mass ratio sBHs in an AGN disc, or generally any encounter, is far from certain to result in a merger, even at a trap location. This suppression of mergers could explain why no event with a mass ratio as extreme as GW190814 has been observed since the detection. 

The binary captures and mergers we have observed in our simulations generally have a small but non-negligable eccentricity of $e \approx 0.01$ when entering the LIGO band at $f_{\rm GW}=10 \rm{Hz}$ (Fig. \ref{fig:orb_evo_merger}). It is unlikely that this eccentricity is detectable with current detectors (consistent with the lack of evidence of eccentricity in GW190814), but it may be detectable with A\# detectors \citep{lower2018measuring}. We have not accounted for BH spins in our simulations: We leave a more detailed study on the expected distribution of parameters such as the individual BH spins, and the mass weighted linear combination of the initial spins aligned with the orbital angular momentum ($\chi_{\rm eff}$) to future work.

\subsubsection{Prospects for host galaxy localisation}

Simplified 1D simulations have shown that migrating BHs encounter one another most frequently near a trap location \citep{vaccaro2025role}. Our simulations indicate that for various discs, binary capture and merger probability at trap locations is a strong function of the central SMBH mass and accretion rate. Thus, the input spectrum of BH masses residing in the AGN disc will be different than the merger mass ratio distribution due to strong modulation by the capture probability (e.g. Figure \ref{fig:q_param_space}). 

Since the AGN parameters can be indirectly inferred from observations of host galaxy mass and luminosity \citep{kormendy2013coevolution,wu2022catalog}, our mass-dependent merger probability could inform a model for weighting potential host galaxies. This will allow informing electromagnetic (EM) follow-up searches for counterpart AGN flares, or cosmological analyses. For quiescent (non-AGN) galaxies, the rate of BBH mergers for a given galaxy scales with the galaxy mass and star formation rate \citep{cao2018host}. Many dark-siren approaches to measuring the Hubble constant from BBH mergers use these scalings \citep{gray2020cosmological}. However, as we have seen, the intricate physical processes of MMR, thermal torques, and gap opening in AGN discs greatly modify the rates of mergers of different mass ratios. Exploring the impact on dark siren analysis from a substantial fraction of mergers originating from the AGN channel is an important avenue for future research. Furthermore, weighting AGN that are the most likely to host BBH mergers can provide extra discriminating power in clustering-based analyses \citep[e.g.,][]{, veronesi2025constraining, Moncrieff2025} to investigate the origins of BBH systems. 

The gas rich environment of the AGN disc potentially allows EM counterparts emitted from kicked remnant BHs interacting with the disc \citep{kimura2021outflow,tagawa2023observable}. The event GW190521 has claimed AGN flare counterparts \citep{graham2020candidate}, and other GW events show tentative evidence \citep{graham2023light} (however see \citealp[e.g.,][for alternatives]{palmese2021ligo, veronesi2025agn}). This great potential for multi-messenger astronomy is currently undermined by significant uncertainties present in the expected EM signature resulting from such a merger \citep{morton2023gw190521}, partially due to lack of knowledge of the merger location and local properties. For an observed BBH merger of masses $m_1$ and $m_2$, a coincident host AGN galaxies EM observability could be partially assessed from the disc properties in the regions where captures are most likely to occur ($\mathscr{B} = \tau_{\rm lib}/\Delta t_{\rm res} \approx 3$). This information could be used to inform the likelihood of a flare being associated with an AGN BH merger versus other scenarios such as tidal disruption events \citep{chan2019tidal} or supernova explosions \citep{gri21}.

\subsection{Caveats and future work}

\textit{Additional bodies and effects:} In this study we mainly limited ourselves to fixed black hole masses $m_1=23.2 M_{\odot}$ and $m_2=2.59 M_{\odot}$. The resonances formed between these masses may be broken or modified by several external sources. Additional bodies in the disc could either destabilize the resonance, or form a resonance chain. Alternatively, resonances at traps may be broken by stochastic torques \citep{secunda19}, due to scattering from additional bodies in a nuclear star cluster and/or turbulent density fluctuations in the disc \citep{trani2025turbulent, epstein-mmr2025}. Additionally, some of the high $k$ resonances we have observed are stabilised from rapid eccentricity growth due to eccentricity damping. Once gas torques fade away as the AGN becomes inactive, the resulting eccentricity growth could lead to resonance breaking, potentially leading to a merger.

\textit{AGN disc parameter space:} We have focused on discs that have $m_{\rm gap} > 2.59 M_\odot$. As noted in Sec. \ref{sec:Background}, the gap mass given by Eqn. \ref{eq:gap_opening_mass} is derived by estimating the angular momentum transport via viscous and gravitational torques, and does not include thermal effects. Thus, the estimate is not always accurate as seen in Figures \ref{fig:Mgap_hr} and \ref{fig:type_2_torques_plots}. Moreover, setting discs with $m_{\rm gap} < 2.59M_\odot$ is challenging since they often have discontinuities in the opacity and hence disc temperature as a function of radius. These jumps affect the estimates for the torques since they sensitively depend on the temperature and density gradients, and can lead to unphysical migration trap locations. Thus, we used the four disc models with $m_{\rm gap} > 2.59 M_{\odot}$. Future work on the population study of mergers in AGN discs will require addressing this issue, probably with smoothing for several length cells.

\textit{Hydrodynamics and circumbinary discs:} Although we've added some prescriptions of gas dynamical friction and eccentricity and inclination damping, our treatment of a bound binary is incomplete. First, the overall radial torque is no longer correct and needs to be applied on the centre of mass of the binary with the total mass. Nevertheless, the timescales associated with migration are much longer than the typical merger time once a binary has formed. The other issue is that we do not account for any feedback on the AGN discs. Generally, a circumbinary accretion disc may form \citep{artymowicz1994dynamics}, which may further harden the binary and drive it to a tight eccentric system \citep{grobner2020binary}. This occurs on timescales comparable to the accretion timescale, which could also be too long for our N-body simulations. Currently most of the hydrodynamical simulations utilise the shearing box approximation (e.g., \citealp{li2022hydrodynamical}). While global simulations of static, narrow annuli are integrated for about $100\ \rm yr$ \citep[e.g.,][]{row23}, capture into and out of MMR require coupled N-body and AGN disc models integrated for long term timescale which is comparable with the migration and libration width to further test our prescriptions. Updates to the capture dynamics from inclusion of additional hydrodynamical effects could lead to a larger value of the free parameter $\zeta$ in Eqn. \ref{eq:kmax}, with captures occuring more frequently than the current fit would suggest. Our choice of $\zeta=1$ does not change any of our result, only changing the fit parameter $\mu$, resulting in the capture probability peaking at a slightly different value of $\mathscr{B}$. 

\textit{Accretion onto the BH:} Accretion onto the BH can alter its mass (and associated torque) and change the dynamics \citep{yi2018growth}. Although for Eddington limited accretion, the mass doubling timescale $\gtrsim 10^8\ \rm Myr$ is usually longer than the AGN lifetime, several studies have proposed that massive BH mergers can form in AGN discs due to accretion \citep{bartos2025accretion}. Accretion is also invoked by \cite{yang2020black} to explain the small BH mass $m_2=2.59 M_{\odot}$ in GW190814. We have not considered the impact of accretion in this study, but since our work is concerned with encounter outcomes of BHs with masses $m_1$ and $m_2$, accretion is negligable on the timescale of this encounter, and therefore can be safely ignored. However, accretion is important for assessing the feasibility of such such an encounter between $m_1$ and $m_2$ occuring in the disc, and thus is an important consideration for future studies. 

\textit{Extreme Mass Ratio Inspirals:}  Black holes in AGN discs are also a key source of extreme mass ratio inspirals (EMRIs, \citealp{pan2021formation}), when a stellar mass black hole plunges into the central supermassive black hole. This will be detectable with space based GW interferometers such as LISA \citep{amaro2023astrophysics}. We have found in our simulations that the resonances formed from encounters at the inner-most trap locations can migrate to the central SMBH within the age of a typical AGN lifetime, provided $M_{\rm SMBH} \lesssim 10^7 M_{\odot}$, leading to the formation of ``double EMRIs'' (similar to \citealp{peng2023synchronizing} and \citealp{peng2024fate}). We will explore this scenario in more detail in a future work.

\textit{Empirical prescriptions and full population study:} We defer a further population study with many black holes within the disc (coming from a realistic distribution of BH masses) to future work. This would be neccesary to get accurate estimation of both BBH merger rates from AGN for ground based detectors, as well as EMRI rates for space based detectors.

However, we have shown that simple analytic models can be can be easily applied to existing Monte-Carlo and 1D simulations for population studies of merger rates and parameter distributions (e.g., \citealp{tagawa2020formation,mckernan2024mcfacts,vaccaro2025role}), giving a reasonable proxy for the dynamics of resonances found with our N-body simulations.

\section{SUMMARY}
\label{sec:conclusion}

In this paper we have explored the evolution of unequal mass black holes (BH) in a variety of AGN discs. We used \texttt{pAGN} \citep{pAGN-code} to generate the AGN disc models and the associated mass-dependent torques for each BH mass, which is self-consistently loaded to \texttt{TSUNAMI} \citep{tsunami-code}. We've accompanied our numerical exploration with semi-analytical modelling of the overall behaviour of BHs in AGN discs and the gravitational-wave merger probability. Our conclusions can be summarised as follows:

$\bullet$ We find three common outcomes described in Sec. \ref{sec:Results}: i) Binary capture and merger (Figure \ref{fig:orb_evo_merger}), ii) capture into high order mean motion resonance (MMR) without close encounter (Figure \ref{fig:resonance_close_enc}), and iii) orbit crossing without the possibility of future close encounters.

$\bullet$ After orbit crossing, several pathways exist for further evolution: Both bodies can park at different traps with no future interaction, or the heavier mass spirals into the SMBH as an EMRI without additional traps. Alternatively, divergent migration can result in the formation of a low order MMR "trap" where there is no net migration, even when gas torques are non-zero for both objects (Figure \ref{fig:resonance_trap_orbit}).

$\bullet$ An additional outcome is a iv) $1:1$ resonant `tadpole' orbit, which is unstable and could lead either to orbit crossing (e.g. Figure \ref{fig:tad_pass}) or binary capture and merger (e.g. Figure \ref{fig:tadpole_to_merger}). 

$\bullet$ We find that a single parameter $\mathscr{B}$, the ratio between the libration timescale and the resonance crossing timescale \citep[see Eqn. \ref{eq:resonance_capture_con} and ][]{batygin2015capture} largely determines the outcome. Orbit crossing and high order MMR occur for $\mathscr{B} \gtrsim 1$ and $\mathscr{B} \ll 1$, respectively, while mergers and tadpoles occur when $\mathscr{B}$ is of order unity (Figure \ref{fig:flowchart}).

$\bullet$ We use a large suite of N-body simulations and fit the capture probability using a log-normal distribution with the fitting parameters reported in Figure \ref{fig:merger_fit}. We use this fit for demonstrating its robustness for various AGN discs and BH mass parameters (Figures \ref{fig:outcome_fractions_discs}-\ref{fig:q_param_space}), applicable for future population studies.

$\bullet$ For our limited parameter space, we tentatively find that the extreme mass ratio compatible with GW190814 merger is optimal for the lower luminosity AGN disc models ($L_{\rm AGN} \approx 10^{43.5}\ \rm erg\ s^{-1}$, Figure \ref{fig:q-L_plot}), however a much more systematic exploration will be carried out in future work. 

Our results serve as an upper limit since some of the tadpoles do not result in mergers. Additional effects such as gas-assisted capture due to individual discs \citep{row23} could increase the capture probabilities. Other effects such as many body interaction \citep{secunda19} or stochastic torques \citep{trani2025turbulent, epstein-mmr2025} could disrupt the MMRs and lead to chaotic encounters that lead to additional mergers. The coupled evolution of these competing effect will be explored in future work.

\section*{Acknowledgements}

We are grateful to Marguerite Epstein-Martin, Shmuel Gilbaum, Nicholas C. Stone, Davide Gerosa, Alberto Sesana, Mor Rozner and Maria Paola Vaccaro for helpful discussions. JWNM, EG and FHP acknowledge support from the OzGrav Research and Innovation Grant. EG acknowledges support from the ARC Discovery Program DP240103174 (PI: Heger). AAT acknowledges support from the Horizon Europe research and innovation programs under the Marie Sk\l{}odowska-Curie grant agreement no. 101103134. FHP is supported by a Forrest Research Foundation Fellowship. JWNM is supported by funding from
the Australian Government Research Training Program. The authors
are grateful for computational resources provided by the OzSTAR
Australian national facility at Swinburne University of Technology.

\section*{Data Availability}
The data underlying this article will be provided by the corresponding author upon reasonable request.



\bibliographystyle{mnras}
\bibliography{bibtemplate}


\bsp	
\label{lastpage}
\end{document}